\documentclass{article}
\usepackage{timet,color}
\usepackage[urlcolor=blue,citecolor=black,linkcolor=black]{hyperref}
\usepackage{natbib}
\usepackage{amsmath,amssymb,amsfonts}
\usepackage[]{graphicx}
\usepackage[pagewise,displaymath, mathlines]{lineno}
\usepackage{authblk}
\usepackage[margin=25mm]{geometry}
\newcommand{\argmax}{\mathop{\rm{arg~max}}\limits}

\newcommand{\ReviewerA}{black}

\begin{document}

\title{Observation Site Selection for Physical Model Parameter Estimation toward Process-Driven Seismic Wavefield Reconstruction}
\author[1,2]{K. Nakai\footnote{Corresponding to kumi.nakai@aist.go.jp}}
\author[1]{T. Nagata}
\author[1]{K. Yamada}
\author[1]{Y. Saito}
\author[1]{T. Nonomura}
\author[3]{M. Kano}
\author[4]{S. Ito}
\author[4]{H. Nagao}
\affil[1]{Graduate School of Engineering, Tohoku University}
\affil[2]{National Institute of Advanced Industrial Science and Technology}
\affil[3]{Graduate School of Science, Tohoku University}
\affil[4]{Earthquake Research Institute, The University of Tokyo}
  

\let\leqslant=\leq

\newtheorem{theorem}{Theorem}[section]

\label{firstpage}

\maketitle

\renewcommand{\thefootnote}{}
\footnote[0]{This paper has been accepted for publication in the Geophysical Journal International \copyright 2023. This manuscript version is made available under the CC-BY-NC-ND 4.0 license (https://creativecommons.org/licenses/by-nc-nd/4.0/)}

\begin{abstract}
The \color{\ReviewerA}``big'' \color{black}seismic data not only acquired by seismometers but also acquired by vibrometers installed in buildings and infrastructure and accelerometers installed in smartphones will be certainly utilized for seismic research in the near future. Since it is impractical to utilize all the seismic big data in terms of the computational cost, methods which can select observation sites depending on the purpose are indispensable. 
We propose an observation site selection method for the accurate reconstruction of the seismic wavefield by process-driven approaches.
The proposed method selects observation sites suitable for accurately estimating physical model parameters such as subsurface structures and source information to be input into a numerical simulation of the seismic wavefield. The seismic wavefield is reconstructed by the numerical simulation using the parameters estimated based on the observed signals at only observation sites selected by the proposed method. 
The observation site selection in the proposed method is based on the sensitivity of each observation site candidate to the physical model parameters;
the matrix corresponding to the sensitivity is constructed by approximately calculating the derivatives based on the simulations, and then, observation sites are selected by evaluating the quantity of the sensitivity matrix based on the D-optimality criterion proposed in the optimal design of experiments. 
In the present study, physical knowledge on the sensitivity to the parameters such as seismic velocity, layer thickness, and hypocenter location was obtained by investigating the characteristics of the sensitivity matrix. Furthermore, the effectiveness of the proposed method was shown by verifying the accuracy of seismic wavefield reconstruction using the observation sites selected by the proposed method.
\end{abstract}


\section{Introduction}
Estimation of ground motion due to an earthquake is one of the indispensable technologies for the prevention and mitigation of disasters. Accurate and rapid estimation of ground motion facilitates the evaluation of seismic damage to infrastructure such as tall buildings, water and gas pipelines, and power plants. It allows us to conduct effective post-disaster measures and rescues when an earthquake occurs. 
Damage to the infrastructure is sometimes evaluated by analyzing the seismic response of structures due to the ground motion. Detecting the damage to individual structures by installing sensors on each structure in a city is an ideal system; however, it is not realistic because of the expensive implementation. Therefore, a method which estimates the seismic damage to infrastructure using observations of ground motions is needed, where the observations are obtained at observation sites located every $10^{3}-10^{4}$ (m) intervals. Hence, our problem can be formulated as seismic wavefield estimation and reconstruction, that recovers the seismic wavefields utilizing seismograms recorded by an array of seismometers distributed more sparsely than the structures. 

A lot of estimation and reconstruction methods have been proposed by data-driven and process-driven approaches. 
In the data-driven approach, ground motions are traditionally calculated by spatially interpolating ground motions recorded by seismometers \citep{vanmarcke1991conditioned,kameda1994conditioned,kameda1992interpolating,sato1999real,kawakami1989simulation}. The interpolation methods utilize nonstochastic conditional simulations under the constraint of given deterministic motions and coherency \citep{kawakami1989simulation}, closed-form solutions of conditional probability functions for Fourier coefficients \citep{kameda1992interpolating},
and regressive (Kriging) models \citep{vanmarcke1991conditioned}, that have been empirically derived based on the relation between seismograms recorded by seismometers and characteristics physical quantities. More recently, the seismic-wave gradiometry (SWG) method has been proposed to reduce the constraint required in the reconstruction \citep[e.g.,][]{langston2007spatial,langston2007wave,liang2009wave,maeda2016reconstruction,shiina2021optimum}. In the SWG method, the amplitudes of seismic waves and their spatial gradients at an arbitrary point can be interpolated from the observed amplitudes at surrounding stations without making assumptions concerning velocity structures and locations of earthquake epicenters. Furthermore, the reconstruction methods utilizing the idea of compressed sensing have been proposed: a method using the sparsity of dominating energies of phase arrivals in the frequency-wavenumber domain \citep{schneider2018improvement}, that using a sparse representation of the surface wavefield using a plane-wave basis \citep{zhan2018application}, and a split processing scheme based on a wavelet transform in time and pre-conditioned curvelet-based compressed sensing in space \citep{muir2021seismic}. The software has been developed for rapid estimation of the seismic damage in the data-driven approach; the ShakeMaps developed by the United States Geological Survey are based on relationships between recorded ground-motion parameters and expected shaking intensities \citep[e.g.,][]{wald1999trinet,wald2005shakemap}. However, the data-driven approaches often suffer from a lack of observation sites and seismograms database. 

While the data-driven approach is based on observations and empirical equations, the wave equations are numerically solved using physical models in the process-driven approach \cite[e.g.,][]{boore1972finite,aoi19993d,pitarka19993d,hisada2003theoretical,koketsu2004finite,ichimura2007earthquake}. The numerical simulations that consider the physics of wave propagation are employed by modelling source mechanisms and subsurface structures. The seismic wavefield and responses estimated by the process-driven approach would be the most reliable if the assumed models were accurate. 
A quick disaster estimation system for the prediction of seismic hazards is developed in the process-driven approach; the system which combines measured ground motion and simulations has been developed by \citet{fujita2014quick}. However, the process-driven approach often involves a high computational cost and requires a high-fidelity subsurface structure model, which is difficult to estimate from observations at the surface alone. 

On the other hand, a combination of process-driven and data-driven approaches has been proposed. \citet{kano2017seismic,kano2017seismicA} proposed a seismic wavefield reconstruction framework by combining seismograms obtained by an array of seismometers and process-based simulations that solve the wave equation. It estimates the physical model parameters related to the subsurface structure and source information by utilizing the replica-exchange Monte Carlo method (REMC) \citep{swendsen1986replica,geyer1991markov,hukushima1996exchange,earl2005parallel}. The wavefield is reconstructed using the estimated parameters that quantitatively explain the observations. 
\color{\ReviewerA}In addition, for such inversion problems, which involve finding the values of model parameters that minimise a misfit function between the observed data and the data simulated with the model parameters, \citet{arnold2018interrogation} introduced the interrogation theory. The interrogation theory combines inverse theory, decision theory, and the theory of experimental design to optimize scientific investigations so as to find information that best answers scientific questions of interest. \citet{zhang2022interrogating,zhao2022interrogating} applied the interrogation theory to seismic tomography to estimate the shape, area, or volume of a subsurface structure and demonstrated the effectiveness of the theory. 
\color{black}
However, in order to use the combination method for real-time reconstruction and quick disaster estimation, it is necessary to reduce the computational cost, since the estimation of model parameters is time-consuming. 

Optimisation of observation sites is one of the key technologies for the development of seismology. In Japan, more than 2,000 seismometers have been in operation for more than 20 years. In hazardous regions for severe seismic damage by a large earthquake, a dense seismic network is often established (e.g. Metropolitan Seismic Observation network (MeSO-net) in Tokyo metropolitan area of Japan \citep{hirata2009outline,sakai2009distribution}). The seismic data not only acquired by seismometers but also acquired by vibrometers installed in buildings and infrastructure and accelerometers built into smartphones will be certainly utilized for seismic research in the near future. In that case, the amount of available data will increase by orders of magnitude compared to the current amount of data. It is impractical to use all the seismic big data in terms of the computational cost. Therefore, prior to the advent of the seismic big data era, it is indispensable to develop an observation site selection method which selects data to be employed depending on the purpose. 
Toward the seismic wave field reconstruction, the methods of observation site selection can propose the subsets of observation sites suitable to estimate the seismic wavefield with sufficient accuracy for damage prediction. Furthermore, the reconstruction accuracy of the seismic wavefield might be improved by the observation site selection which considers the characteristics of the signal-to-noise ratio of each seismometer. In addition, when seismometers are newly installed, the observation site selection method can indicate the configuration of seismometers that contributes most to the improvement of accuracy of the seismic wavefield reconstruction. 
Previous studies have attempted to optimise the design of the seismic network configuration. \citet{hardt1994design} developed a method to design optimum networks for aftershock recordings based on simulated annealing \citep{kirkpatrick1983optimization}. This method has been extended in \citet{kraft2013optimization}. \citet{steinberg2003optimal} applied the statistical theory of optimal design of experiments \citep{atkinson2007optimum} to derive network configurations that maximize the precision of earthquake source localisation. More recently, \citet{muir2022wavefield} has investigated an optimal design of mixed distributed acoustic sensing (DAS), a combined network of DAS and point sensors, based on the optimal design of experiments.
\color{\ReviewerA}Furthermore, for the seismic source inversion, \citet{long2015fast} proposed a technique based on the optimal design of experiments in a Bayesian framework \citep{chaloner2015bayesian,stuart2010inverse}. They proposed the expected information gain obtained from the data recorded by an array of receivers (seismographs), which corresponds to the optimality criterion in the optimal design of experiments, and a fast estimator of the expected information gain by employing the Laplace approximation. In the numerical experiments assuming equally spaced receivers, the distance between the receivers to maximise the gain has been investigated. \color{black}

This type of of situation, where the quantities of physical parameters are estimated based on observed data using discretely installed sensors such as seismographs, can be seen in a range of scientific fields. It is valuable to select the locations to install sensors from the candidate locations in order to obtain as much useful information as possible with as few sensors as possible. This sensor selection/placement problem, which is called the sparse sensor optimisation problem, are intensively studied for the global positioning system \citep{kihara1984satellite,phatak2001recursive}, structural health monitoring \citep{worden2001optimal,yi2011optimal}, environmental monitoring \citep{du2014optimal}, brain source localisation \citep{yeo2022efficient}, and flow visualization \citep{manohar2018data,carter2021data,kanda2021feasibility,kanda2022proof,kaneko2021data,inoue2021data,inoue2023data,inoba2022optimization,tiwari2022simultaneous,li2021data,li2021efficient,fukami2021global,callaham2019robust}, etc. 
The sparse sensor optimisation methods have been proposed using various schemes such as a convex relaxation method \citep{joshi2009sensor,liu2016sensor,nonomura2021randomized,nagata2021data,nagata2022data}, and a greedy method \citep{manohar2018data,manohar2021optimal,manohar2019optimized,clark2018greedy,clark2020multi,clark2020sensor,saito2020data,yamada2021fast,nakai2021effect,nakai2022nondominated,nagata2023randomized,yamada2022greedy}. 

The present study proposes an observation site selection method for the accurate reconstruction of the seismic wavefield from sparse observation. The framework of sparse sensor optimisation based on linear observation equations is extended to seismic wavefield reconstruction. The observation site selection method proposed in the present study is for the process-driven reconstruction method; the locations of observation sites are optimised to estimate the physical model parameters based on a linear equation which expresses the relation between the parameters and the sparse observations, and the seismic wavefield is reconstructed using the simulation with parameters estimated by the observations at the optimised observation sites. 
The observation site selection method proposed in the present study is expected to contribute to improving the rapidness of the process-driven seismic wavefield reconstruction, e.g., the method proposed in \citet{kano2017seismic}. In the proposed reconstruction framework, the model parameters are estimated using the replica-exchange Monte Carlo method, which allows a parameter search in a wide parameter space but is time-consuming. 

The remainder of this paper is organized as follows: Section~\ref{sec:methodologies} formulates an observation site selection method for seismic model parameter estimation toward accurate seismic wavefield reconstruction. Section~\ref{sec:selection} presents the results of observation site selection using the proposed method. Section~\ref{sec:twinexp} demonstrates the improvement in the reconstruction accuracy of the numerically simulated seismic wavefield based on the data at the observation sites selected by the proposed method compared to that at observation sites randomly selected. Section~\ref{sec:conclusions}  concludes the paper.

\section{Methodologies}
\label{sec:methodologies}

\subsection{Sparse sensor optimisation}
\label{sec:methodologies_sparse}
The formulation and algorithms of sparse sensor optimisation proposed in the field of fluid dynamics are described in this subsection \citep{manohar2018data,saito2021determinant,saito2021data}. The observation of ground motions using seismometers corresponds to the vector-sensor measurement. The acceleration is typically measured in the north-south, east-west and up-down directions at every single observation site. The number of multiple components is further increased when considering the seismic waveform in the frequency domain. 
We define the vector-measurement-sensor optimisation problem as follows:
\begin{eqnarray}
\mathbf{y}
    &=& \left[\begin{array}{c}
            \mathbf{H}_{s_{1}}\\
            \mathbf{H}_{s_{2}}\\
            \vdots\\
            \mathbf{H}_{s_{p}} \\
        \end{array}\right]
        \left[\begin{array}{c}
            \mathbf{U}_1 \\ \mathbf{U}_2 \\ \vdots \\ \mathbf{U}_s 
        \end{array}\right] 
        \mathbf{z}\nonumber\\
    &=& \left[\begin{array}{c}
            \mathbf{W}_{1}\\
            \mathbf{W}_{2}\\
            \vdots\\
            \mathbf{W}_{p}\\
        \end{array}\right]
        \mathbf{z}\nonumber\\
    &=& \mathbf{D} \mathbf{z}\label{eq:linearobs_vec}
\end{eqnarray}
where $\mathbf{y} \in \mathbb{R}^{p} $ is the observation vector, $\mathbf{H}_{s_{k}} \in \mathbb{R}^{s \times N}$ is the sensor location matrix of $k$th sensor for first-to-$s$~th vector components, $\mathbf{U}_s \in \mathbb{R}^{\frac{N}{s}\times r}$ is the $s$th vector component of a sensor candidate matrix, and $\mathbf{z} \in \mathbb{R}^{r}$ is the latent variable vector. Here, $p$, $r$, $s$ and $N$ are the number of sensors to be selected, the number of latent variables, the number of components of the measurement vector, and the number of spatial dimensions including the different vector components, respectively. 
Let $\mathbf{W}_{k} \in \mathbb{R}^{s\times r}$ denotes the $k$th vector-sensor-candidate matrix as follows:
\begin{align}
    \mathbf{W}_{k} = \left[\begin{array}{cccc}
        \mathbf{u}_{i_{k},1}^{\mathrm{T}} & \mathbf{u}_{i_{k},2}^{\mathrm{T}} & \dots & \mathbf{u}_{i_{k},{s}}^{\mathrm{T}} \end{array}\right]^{\mathrm{T}},
    \label{eq:Wk}
\end{align}
where $i_k$ and $\mathbf{u}_{i_{k},l} \in \mathbb{R}^{1 \times r}$ are the index of the $k$th selected sensor and the corresponding row vector of the $l$th vector component of the sensor-candidate matrix, respectively.
Let $\mathbf{D} \in \mathbb{R}^{s \times r}$ denotes the vector-sensor-candidate matrix as follows:
\begin{align}
    \mathbf{D} = \left[\begin{array}{cccc}
        \mathbf{W}_{1}^{\mathrm{T}} & \mathbf{W}_{2}^{\mathrm{T}} & \dots & \mathbf{W}_{p}^{\mathrm{T}}
    \end{array}\right]^{\mathrm{T}}.
    \label{eq:D}
\end{align}
The best estimation of latent variables $\tilde{\mathbf{z}}$ can be obtained by the pseudo-inverse operation when uniform independent Gaussian noise is imposed on the observations. As described in \cite{saito2021determinant,nakai2021effect}, the objective function for sensor optimisation problems in the linear estimation can be defined using the Fisher information matrix (FIM), which corresponds to the inverse of the covariance matrix of estimation error, based on the optimal design of experiments \citep{atkinson2007optimum}. The most commonly used criterion is the D-optimality criterion, which is equivalent to minimizing the determinant of the error covariance matrix, whereas there are a variety of criteria proposed in the optimal design of experiments. The vector-measurement-sensor optimisation based on the D-optimality criterion can be expressed as the optimisation problem as follows \citep{saito2021data}:
\begin{align}
    &\mathrm{maximize}\,\,f_{\mathrm{D}_{s}} \nonumber \\
    &f_{\mathrm{D}_{s}}\,=\,\left\{\begin{array}{cc}
        \mathrm{det}\,\left(\mathbf{D}\mathbf{D}^{\top}\right), & p\le r/s, \\
        \mathrm{det}\,\left(\mathbf{D}^{\top}\mathbf{D}\right), & p>r/s.
    \end{array}\right.\label{eq:obj_detvec}
\end{align}

The sensor optimisation problem is formulated as a combinatorial optimisation problem, which is known as an NP-hard problem. Greedy methods have been devised and applied to the sensor optimisation problems instead of a brute-force algorithm, which evaluates all the combinations of sensors out of sensor candidates and takes enormous computational cost. In the step-by-step selection using the greedy method, only the $k$th sensor selection is carried out in the $k$th step under the condition that the first-to-($k-1$)th sensors are already determined. 

Furthermore, a unified expression for the undersampled and oversampled cases has been proposed in \citet{saito2021determinant,saito2021data}, whereas different objective functions are defined in the undersampled and oversampled cases in which the number of sensors is less than and greater than that of the latent state variables, respectively. The function $\mathrm{det}\left(\mathbf{D}^{\top}\mathbf{D}+\epsilon\mathbf{I}\right)$ has been proposed as the approximated objective function, where $\epsilon$ is a sufficiently small number \citep{saito2021determinant,shamaiah2010greedy}. The sensor index chosen in $k$th step of the greedy algorithm can be described as follows:
\begin{align}
    i_{k}=&\argmax_{i_k}\,\,\mathrm{det}\,\left(\mathbf{D}_k^{\top}\mathbf{D}_k+\epsilon\mathbf{I}\right). \label{eq:obj_detvec2} 
\end{align}

Although eq.~\eqref{eq:obj_detvec} was adopted in the initial phase of the present study, this objective function was found to poorly work for the situation in which vector sensors in one sensor location are not linearly independent. This is because a vector sensor in one sensor location has more than 1230 components in the present situation as later discussed and eq.~\eqref{eq:obj_detvec} for the $p>r/s$ condition should be used even in the selection of the first sensor, but it gives rank-deficient FIM which corresponding to $\mathrm{det}\,\left(\mathbf{D}^{\top}\mathbf{D}\right)=0$ for any sensor selected for the first step. Therefore, the greedy method in eq.~\eqref{eq:obj_detvec} does not work well for the first step. This situation is relaxed by using the unified and robust formulation owing to the regularization of FIM though the hyperparameter $\epsilon$ should be introduced. 
Therefore, we employ the formulation \eqref{eq:obj_detvec2} for the vector-sensor-measurement-optimisation problem in the present study.

The greedy method based on the objective function \eqref{eq:obj_detvec2} is confirmed to provide the same sensors as the convex relaxation method using eq.~\eqref{eq:obj_detvec2} \citep{saito2021data}, which obtains a nearly optimal solution. Furthermore, the objective function based on the A-optimality criterion, which also has a statistical interpretation in terms of FIM as with D-optimality criterion and evaluates the trace of the inverse of FIM, can be defined in a similar way to D-optimality criterion \citep{nakai2021effect}. The results using the objective function based on the A-optimality criterion are confirmed to be almost the same as those based on the D-optimality criterion. Therefore, the results based on A-optimality criterion are omitted for brevity in the present paper. 

\subsection{Observation site selection based on sensitivity to model parameters}
\label{sec:methodologies_selection}
In the present study, an observation site selection method which accurately estimates the model parameters used as inputs in the numerical simulation of the wave field is proposed, whereas the seismic wavefield is assumed to be reconstructed by a numerical simulation. 
Throughout this study, we use the code developed by \citet{hisada1995efficient} for the numerical calculation. This code calculates the theoretical waveforms at a certain distance from the epicenter location based on the discrete wavenumber method, in which seismic waveforms are represented as wavenumber integrals of Green's functions. Theoretical waveforms are obtained by assuming a one-dimensional (1-D) horizontally layered subsurface structure model and given source information. 

\begin{figure}
    \centering
    \includegraphics[width=4cm]{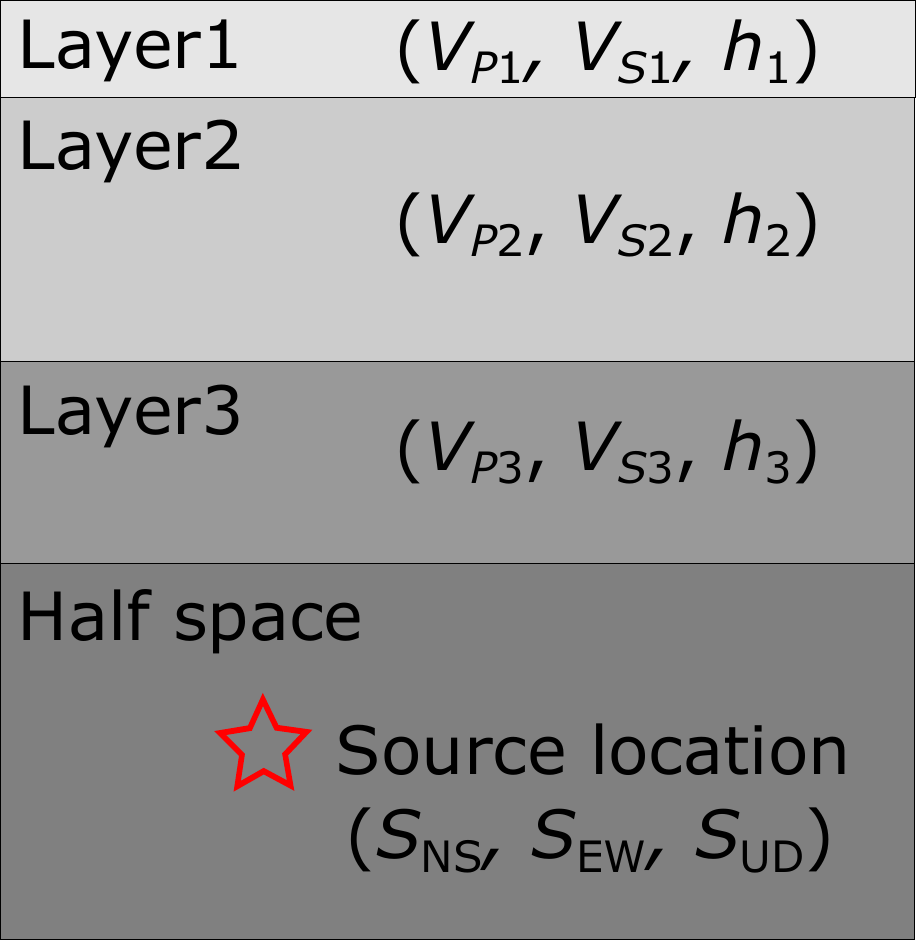}
    \caption{Multi-layered subsurface structure model consisting of three layers on a half-space. The settings of the subsurface structure model parameters ($V_P$, $V_S$, and $h$) and the source location parameters ($S_{\rm{NS}}$, $S_{\rm{EW}}$, and $S_{\rm{UD}}$) are mentioned in Tables~\ref{tab:ModelParam} and \ref{tab:HypoParam}, respectively.}
    \label{fig:3layers}
\end{figure}

Figure~\ref{fig:3layers} shows a multi-layered subsurface structure model assumed in the present experiment.
Therefore, the seismic velocities ($V_P$ and $V_S$) and the thickness ($h$) of each layer are the target model parameters to be estimated. In addition, earthquake source location and fault parameters such as strike, rake, and dip angles are the physical parameters to estimate.  

The waveform data observed at each observation site is obtained as Fourier spectra using the code developed by \citet{hisada1995efficient}; it has $i$ frequency components, and it has six components for each frequency component: real and imaginary parts every three components for three directions (north-south (NS), east-west (EW), and up-down (UD)). Hence, the Fourier coefficients obtained at $n$th observation site are stored in a column vector $\mathbf{x}_n\in\mathbb{R}^{(6\times i)\times1}$ in the proposed method as follows: 
\begin{equation}
    \mathbf{x}_n
    =
    \left[\begin{array}{ccccccccc}
        \text{Re}(\xi^1_1) & \text{Im}(\xi^1_1) & \text{Re}(\eta^1_1) & \text{Im}(\eta^1_1) & \text{Re}(\zeta^1_1) & \text{Im}(\zeta^1_1)& \text{Re}(\xi^2_1) & \dots & \text{Im}(\zeta^i_1)
    \end{array} \right]^{\mathrm{T}}, \label{eq:xvec_oneobs}
\end{equation}
where $\xi$, $\eta$, and $\zeta$ are the components of a certain frequency of Fourier spectra for three directions in space (NS, EW, UD), respectively. The waveform data matrix shown in eq.~\eqref{eq:xvec_oneobs} is obtained at all observation sites. Thus, the waveform data vector, which is constructed by stacking the data for all observation sites $\mathbf{X}\in\mathbb{R}^{(6\times i\times n)\times1}$ is following form.
\begin{equation}
    \mathbf{X}=
    \left[  
    \begin{array}{cccc}
    \mathbf{x}_1^{\mathrm{T}} & \mathbf{x}_2^{\mathrm{T}} & \dots & \mathbf{x}_n^{\mathrm{T}}
    \end{array}
    \right]^{\mathrm{T}} \label{xvec_allobs}
\end{equation}

We consider the linear equation which describes the relation between the parameters to be estimated and observation data at observation sites since the framework proposed assumes a linear observation equation as written in eq.~\eqref{eq:linearobs_vec}:
\begin{equation}
    \left[
    \begin{array}{c}
    \mathbf{\delta x}_1 \\ \mathbf{\delta x}_2  \\ \vdots \\ \mathbf{\delta x}_n
    \end{array}
    \right]    
    =
    \left[
    \begin{array}{cccc}
     \tilde{\phi}_1\frac{\partial \mathbf{x}_1}{\partial \phi_1} & \tilde{\phi}_2 \frac{\partial \mathbf{x}_1}{\partial \phi_2} & \dots &  \tilde{\phi}_r \frac{\partial \mathbf{x}_1}{\partial \phi_r} \\
    \tilde{\phi}_1\frac{\partial \mathbf{x}_2}{\partial \phi_1} & \tilde{\phi}_2 \frac{\partial \mathbf{x}_2}{\partial \phi_2} & \dots & \tilde{\phi}_r \frac{\partial \mathbf{x}_2}{\partial \phi_r} \\
    \vdots & \vdots & \vdots & \vdots \\
    \tilde{\phi}_1\frac{\partial \mathbf{x}_n}{\partial \phi_1} & \tilde{\phi}_2\frac{\partial \mathbf{x}_n}{\partial \phi_2} & \dots & \tilde{\phi}_r \frac{\partial \mathbf{x}_n}{\partial \phi_r} \\
    \end{array}
    \right]
    \left[
    \begin{array}{c}
    \frac{\delta \phi_1}{\tilde{\phi}_1} \\ \frac{\delta \phi_2}{\tilde{\phi}_2} \\ \vdots \\ \frac{\delta \phi_r}{\tilde{\phi}_r}
    \end{array}
    \right], \label{eq:linearizedmodel}
\end{equation}
where $\phi$ is the model parameter to be estimated, and $r$ is the number of model parameters to be estimated, which corresponds to the latent state variables in eq.~\eqref{eq:linearobs_vec}. 
Although the model parameter estimation should be defined as a nonlinear estimation problem, linearisation around the true values of the model parameters is applied in the present study by assuming small errors between the true and initial values of the model parameters. This is a reasonable assumption since the estimation of the model parameters is performed using preliminary report values, which are expected to be close to the true value as initial values in real problems. 
Equation~\eqref{eq:linearizedmodel} expresses the reconstruction error in the waveform data at each observation site ($\mathbf{\delta x}$) that is caused by the estimation errors of the model parameters ($\delta \phi$);
the second term on the right corresponds to the differences between the true and estimated values of the model parameters, and the left-hand side corresponds to the differences between the observed data and data reconstructed using a numerical simulation with the estimated model parameters. 
It should be noted that the estimation error of each model parameter is normalized by the initial value ($\tilde{\phi}$).
Hence, the first term on the right in eq.~\eqref{eq:linearizedmodel} corresponds to a matrix that represents the normalized sensitivity to the model parameters (hereinafter, referred to as normalized parameter sensitivity matrix); each row of the matrix corresponds to the sensitivity of each observation site candidate to the model parameters. 
Therefore, the observation sites with high sensitivity to the model parameters can be selected by evaluating the quantity of the normalized parameter sensitivity matrix. In the framework of the sparse sensor optimisation described in Section~\ref{sec:methodologies_sparse}, the normalized parameter sensitivity matrix corresponds to the vector-sensor-candidate matrix $\mathbf{D}$ in eq.~\eqref{eq:linearobs_vec}. The D-optimality criterion is adopted in the present study, and the observation sites are selected by the D-optimality-based greedy method using the objective function \eqref{eq:obj_detvec}. \color{\ReviewerA}Note that the observed noise is assumed to follow a normal distribution in the formulation of the sparse sensor optimisation described in Section~\ref{sec:methodologies_sparse}. Thus, the result of the observation site selection by the D-optimality-based greedy method using the objective function eq.~\eqref{eq:obj_detvec} does not depend on the noise following a normal distribution contained in the measurement data.  On the other hand, several researchers have studied sensor selection in the presence of correlated measurement noise \citep{liu2016sensor,ucinski2020d,yamada2021fast,nagata2022data}. The observation site selection method proposed in the present study can be extended to the case of correlated noise by utilizing the formulation con considering correlated measurement noise, which is left for the future study. \color{black}

\subsection{Model parameter estimation method}
\label{sec:methodologies_reconst}
The estimation method of the model parameters based on observation at the selected observation sites by the proposed method and the reconstruction method of the seismic wavefield is explained.
The waveform data vector for the subset of observation sites selected by the proposed method $\mathbf{Y}\in\mathbb{R}^{(6\times i\times p)\times1}$ is described as follows:
\begin{equation}
    \mathbf{Y}=
    \left[  
    \begin{array}{cccc}
    \mathbf{y}_1^{\mathrm{T}} & \mathbf{y}_2^{\mathrm{T}} & \dots & \mathbf{y}_{p}^{\mathrm{T}}
    \end{array}
    \right]^{\mathrm{T}}, \label{yvec}
\end{equation}
where $p$ is the number of observation sites selected. 
The values of model parameters are repeatedly updated until the change in error in the reconstructed wavefield approaches zero, as in Newton's method.
The errors of the reconstructed wavefield are evaluated by comparing the observation at the selected observation sites as follows: 
\begin{equation}
    \left[  
    \begin{array}{c}
    \mathbf{\delta y}_1 \\ \mathbf{\delta y}_2 \\ \vdots \\ \mathbf{\delta y}_{p}
    \end{array}
    \right]    
    =
    \left[  
    \begin{array}{c}
    \mathbf{y}_1  \\ \mathbf{y}_2 \\ \vdots \\ \mathbf{y}_{p}
    \end{array}
    \right]   -
    \left[  
    \begin{array}{c}
    \tilde{\mathbf{y}_1} \\ \tilde{\mathbf{y}_2} \\ \vdots \\ \tilde{\mathbf{y}}_{p}
    \end{array}
    \right].   \label{eq:reconst_seiserr}
\end{equation}
Then, the estimation errors of the model parameters are calculated using the parameter sensitivity matrix consisting of rows corresponding to the selected observation sites and the pseudo-inverse operation as follows:
\begin{equation}
    \left[
    \begin{array}{c}
    \delta \phi_1  \\ \delta \phi_2 \\ \vdots \\ \delta \phi_r
    \end{array}
    \right]
    =
    \left[
    \begin{array}{cccc}
     \frac{\partial \mathbf{y}_1}{\partial \phi_1} & \frac{\partial \mathbf{y}_1}{\partial \phi_2} & \dots &  \frac{\partial \mathbf{y}_1}{\partial \phi_r} \\
    \vdots & \vdots & \vdots & \vdots \\
    \frac{\partial \mathbf{y}_{p}}{\partial \phi_1} & \frac{\partial \mathbf{y}_{p}}{\partial \phi_2} & \dots & \frac{\partial \mathbf{y}_{p}}{\partial \phi_r} \\
    \end{array}
    \right]^{\dagger}
    \left[  
    \begin{array}{c}
    \mathbf{\delta y}_1 \\ \mathbf{\delta y}_2 \\ \vdots \\ \mathbf{\delta y}_{p}
    \end{array}
    \right], \label{eq:reconst_paramerr}
\end{equation}
where $\circ^\dagger$ indicates the Moore--Penrose pseudoinverse.
The values of model parameters are calibrated based on the estimation errors in eq.~\eqref{eq:reconst_paramerr}:
\begin{equation}
    \left[  
    \begin{array}{c}
    \phi_1^{t+1} \\ \phi_2^{t+1} \\ \vdots \\ \phi_r^{t+1}
    \end{array}
    \right]    
    =
    \left[  
    \begin{array}{c}
    \phi_1^{t} \\ \phi_2^{t} \\ \vdots \\ \phi_r^{t}
    \end{array}
    \right]   -
    \left[  
    \begin{array}{c}
    \delta \phi_1 \\ \delta \phi_2 \\ \vdots \\ \delta \phi_r
    \end{array}
    \right], \label{eq:reconst_paramcali}
\end{equation}
where $t$ is the number of optimisation steps. Then, the seismic wavefield is reconstructed using the numerical simulation with the updated model parameters, and the errors of seismic wavefield at the selected observation sites are evaluated as described in eq.~\eqref{eq:reconst_seiserr}. The series of the estimation of the model parameters [eq.~\eqref{eq:reconst_seiserr}--\eqref{eq:reconst_paramcali}] is repeated until the change in the residual errors of the reconstructed seismic wavefield approaches to a sufficiently small value at the selected observation sites.

\color{\ReviewerA}Note that the observation site selection explained in Section~\ref{sec:methodologies_selection}
can be performed each time the physical model parameters are updated in the series of the estimation [eq.~\eqref{eq:reconst_seiserr}--\eqref{eq:reconst_paramcali}]; however, the aim of the present study is to propose a method for the preselection of a prioritised list of observation sites, depending on the areas, magnitudes, source mechanisms and earthquake types (e.g., plate-boundary and inland types) from the point of view of an accurate reconstruction of the seismic wavefield. Reselection of observation sites in the estimation process is outside the scope of the present study. In addition, the reselection of observation sites in the estimation process is unlikely to affect the results of the observation site selection in a situation where the difference between the initial and true parameters is small. \color{black}

Our research group has been developing a seismic wavefield reconstruction method based on the data-driven reduced-order model \citep{nagata2022seismic} besides the observation site selection method proposed in the present study. 
The aim of both studies is to achieve an accurate reconstruction of the seismic wavefield. The critical difference between these two studies is that the present study is the process-driven approach, where \citet{nagata2022seismic} is the data-driven approach. Since the process-driven approach requires accurate estimation of model parameters, the present study proposes a method that selects observation sites suitable for accurate seismic wavefield reconstruction. \citet{kano2017seismic} has proposed the data-driven reconstruction framework using a replica-exchange Monte Carlo method for the nonlinear estimation problem of model parameters. For a quick estimation, it is significant to estimate as accurately as possible using information from as few observation sites as possible. The parameter estimation problem is linearised in the observation site selection method proposed in the present study; however, the sensitive location obtained by the proposed method is expected to be also applicable to the parameter search in the practical nonlinear problem. The observation site selection method proposed in the present study is one of the important elemental technologies for the improvement of accuracy and rapidness of the process-driven reconstruction because it can be adapted to various process-driven reconstruction methods.

\section{Observation site selection}
\label{sec:selection}

\subsection{Problem settings for observation site selection}
\label{sec:selection_probset}
The target area for seismic wavefield reconstruction is the Tokyo metropolitan area of Japan, in which a dense seismological network named MeSO-net has been in operation since 2007. The MeSO-net comprises 296 accelerometers, located at intervals of a few kilometers, that continuously record seismograms at a sampling rate of 200~Hz \cite[e.g.][]{hirata2009outline,sakai2009distribution}.
The target area is a sedimentary basin known as the Kanto basin. The Kanto basin on the bedrock can be approximated by three layers lying on a half-space, as shown in Fig.~\ref{fig:3layers}, which follows the Japan Integrated Velocity Structure Model (JIVSM) \citep{koketsu2011unified}. Hence, the subsurface structure is assumed to be a 1-D horizontally layered subsurface structure consisting of three layers, which model the sedimentary basin. The three layers are lying on bedrock, which is modeled as a half-space. Table~\ref{tab:ModelParam} shows the subsurface structure model parameters. 

\begin{figure*}
    \centering
    \includegraphics[width=14cm]{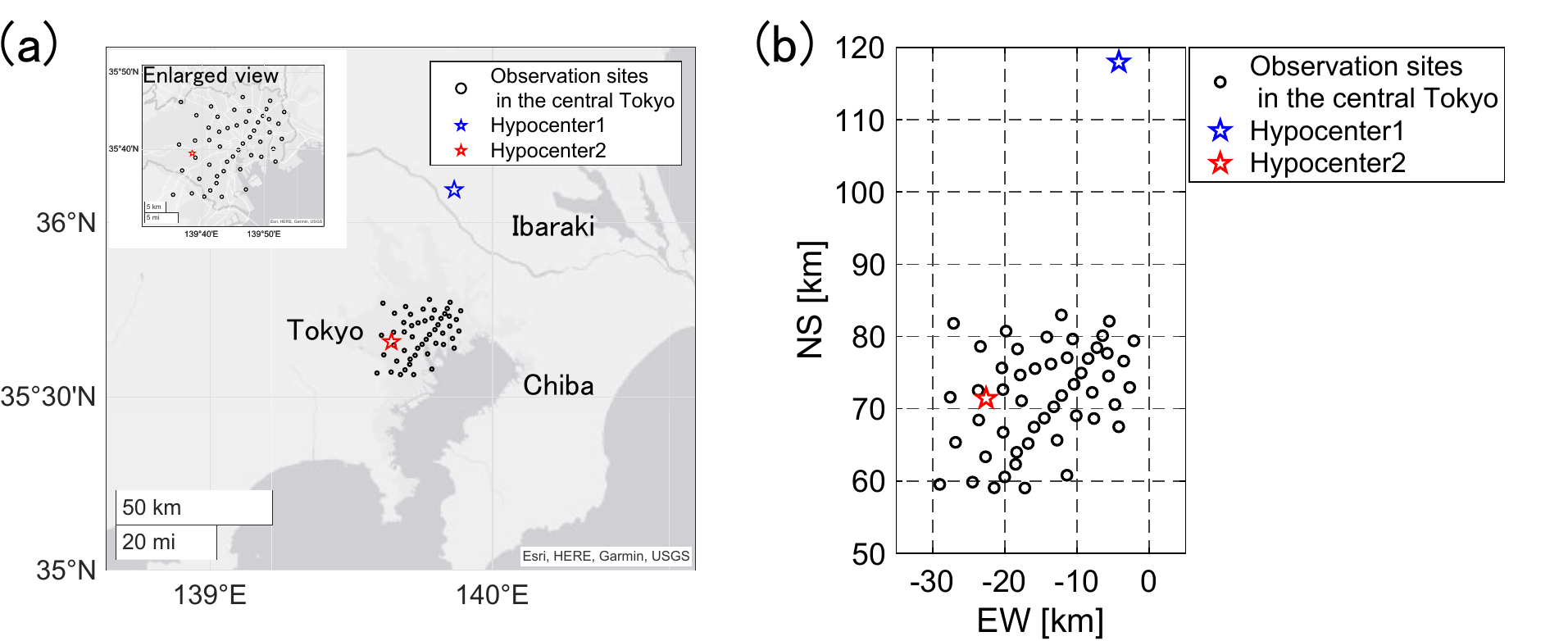}
    \caption{Geometric setting. (a) The Kanto area in the geographic coordinate system. (b) Locations of observation sites and earthquake sources in the north-south(NS)/east-west(EW) coordinate system. The circles indicate the observation sites, which are referred to the actual distribution of the MeSO-net. The blue and red marks indicate the earthquake source, the values of which are mentioned in Table~\ref{tab:HypoParam}. This figure corresponds to the top view of Fig.~\ref{fig:3layers}, and the source is located in the half space as shown in Fig.~\ref{fig:3layers}.}
    \label{fig:obsposi}
\end{figure*}

The observation sites assumed in the present paper are the subset of the observation sites of MeSO-net that actually exist in the central Tokyo of Japan. Figure~\ref{fig:obsposi} shows geometric settings. The circles indicate the assumed observation sites. The number of observation sites is 50. The origin of horizontal (NS-EW) coordinates is set as ($35.0340^{\circ}$N, $139.9106^{\circ}$E). Note that the results do not depend on the setting of the origin of horizontal coordinates since the relative positional relationship between the observation sites and the hypocenter affects the results. 

The earthquake source is assumed to be located in the half-space as shown in Fig.~\ref{fig:3layers}. In the present paper, two types of earthquakes are simulated using different source information, and the characteristics of observation site selection by the proposed method are verified. The source information is set based on the earthquakes actually occurred in the Tokyo metropolitan area. Table~\ref{tab:HypoParam} summarizes the source information. In the case of the hypocenter~1, the source information is set based on that of the earthquake which occurred in the southern part of Ibaraki prefecture on September 16, 2014. This area is known as one of areas where earthquakes occur frequently. In the case of the hypocenter~2, the source information is set based on that of the earthquake which occurred directly underneath the central Tokyo on May 9, 2010. The blue and red marks shown in Fig.~\ref{fig:obsposi} indicate the locations of the hypocenters~1 and 2. The cases of the hypocenters~1 and 2 correspond to situations where the earthquake occurs far from and directly underneath the area of observation sites, respectively. 

\begin{table}
    \caption{Subsurface structure model parameters which are defined as shown in Fig.~\ref{fig:3layers}.}
    \label{tab:ModelParam}
    \begin{tabular}{lllll}
    \hline
     & $\rho$ [g/cm$^3$] & $V_P$ [km/s]& $V_S$ [km/s]& $h$ [km] \\ \hline
    Layer 1 & 1.95 & 1.8 & 0.5 & 0.4 \\
    Layer 2 & 2.15 & 2.4 & 1.0 & 1.1 \\
    Layer 3 & 2.3 & 3.2 & 1.7 & 1.0 \\
    Half space & 2.7 & 5.8 & 3.4 & $\infty$ \\ \hline
    \end{tabular}
\end{table}

\begin{table}
    \caption{Source parameters of two earthquakes; the hypocenters~1 and 2 correspond to situations where the earthquake occurs far from and directly underneath the area of observation sites, respectively, as shown in Fig.~\ref{fig:obsposi}.}
    \label{tab:HypoParam}
    \begin{tabular}{ll}
    \hline
    Hypocenter~1 (Southern part of Ibaraki) & \\ \hline
    Source location (NS, EW, UD) [km] & 117.9655, -4.2204, 47.0000  \\
    Strike, rake, dip [deg] & 254, 118, 28 \\
    Slip [m] & 0.4 \\ \hline
    Hypocenter~2 (Tokyo 23 Ward) & \\ \hline
    Source location (NS, EW, UD) [km] & 71.4339, -22.5905, 26.0000 \\
    Strike, rake, dip [deg] & 126, 103, 80 \\
    Slip [m] & 0.05 \\ \hline
    \end{tabular}
\end{table}

In the present study, the seismic velocities of each of the three layers ($V_{Pi}$ and $V_{Si}$, where $i=1,2,3$) and the thickness ($h_i$) of each of the three layers, and source location in three directions ($S_{\mathrm{NS}}$, $S_{\mathrm{EW}}$, and $S_{\mathrm{UD}}$) are the estimation target of the model parameters following the previous studies \citep{kano2017seismic,kano2017seismicA}. Note that the origin time, which is estimated in the previous studies \citep{kano2017seismic,kano2017seismicA}, is not taken into account as a target parameter since Fourier spectra of the seismic wavefield is evaluated in the proposed method. The total number of model parameters to be estimated is $r=12$.
The sensitivity matrix is constructed by simulating 24 cases using the code developed by \citet{hisada1995efficient}: 2 cases in which the value of one of the 12 parameters is changed by $+\Delta \phi$ and $-\Delta \phi$ from the initial value shown in Tables~\ref{tab:ModelParam} and \ref{tab:HypoParam} are conducted, and then, the difference in Fourier spectra by $2\Delta \phi$ is obtained for each parameter. The values of $\Delta \phi$ for the subsurface structure model parameters ($V_{Pi}$, $V_{Si}$, and $h_i$) are set to $10\%$ of initial values (Table~\ref{tab:ModelParam}), and those for the source location parameters ($S_{\mathrm{NS}}$, $S_{\mathrm{EW}}$, and $S_{\mathrm{UD}}$) are set to 500~m.

The simulation was conducted in the frequency band from DC to 5~Hz. The number of frequency components is $i=205$. Hence, the number of components obtained by vector measurement at one observation site is $s=1230$ considering each frequency component consisting of the real and imaginary parts as shown in eq.~\ref{eq:xvec_oneobs}. The second-order Butterworth filter was applied to the obtained simulation results in order to verify the influence of the frequency band of the seismic wavefield on the observation site selection. The lower cutoff frequency of the bandpass filter is set to be $0.1$ Hz, and the higher cutoff frequency varies from 0.3~Hz to 1.0~Hz at 0.1~Hz intervals.

\subsection{Normalized parameter sensitivity} 
\label{sec:selection_NPS}
In this subsection, normalized parameter sensitivity $\mathcal{S}$ is newly defined, and the characteristics of the parameter sensitivity matrix of interest in the proposed method are investigated. This is a scalar quantity showing the sensitivity of each observation site to each parameter.

We consider the difference in the observation data due to the difference in the $k$th parameter at the $j$th observation site. This corresponds to the numerator of $\tilde{\phi}_k\frac{\partial \mathbf{x}_j}{\partial \phi_k}$ in the normalized parameter sensitivity matrix in eq.~\eqref{eq:linearizedmodel}. 
The difference in the observation data ($\Delta \mathbf{x}_j^{\phi_k}$) is expressed by the difference between the observation data in the case of $\phi_k=\phi_k+\Delta \phi_k$ and that in the case of $\phi_k=\phi_k-\Delta \phi_k$ as follows:
\begin{equation}
    \Delta \mathbf{x}_j^{\phi_k}
    = \mathbf{x}_j|_{\phi_k=\phi_k+\Delta \phi_k} - \mathbf{x}_j|_{\phi_k=\phi_k-\Delta \phi_k}.
\end{equation}
Here, $\Delta \phi$ in the present study was set to be the same value as that for constructing the normalized sensitivity matrix described in Section~\ref{sec:selection_probset}. Since the observation data $\mathbf{x}$ is a vector as shown in eq.~\eqref{eq:xvec_oneobs}, we define the sum of the squares of each component of the difference vector, that is, the square of the $L_2$ norm as the normalized parameter sensitivity to $k$th parameter of $j$th observation site ($\mathcal{S}_j^{\phi_k}$) as follows:
\begin{equation}
    \mathcal{S}_j^{\phi_k} = ||\Delta \mathbf{x}_j^{\phi_k}||^2_2.
\end{equation}
The scalar value $\mathcal{S}_j^{\phi_k}$ can be evaluated for each observation $(j=1,\dots,50)$ and for each parameter $(k=1,\dots,12)$.

It should be noted that the quantity evaluated in the observation site selection is not exactly the same as $\mathcal{S}_j^{\phi_k}$. The proposed site selection method evaluates the value of sites based on the determinant of the normalized parameter sensitivity matrix which is the quantity showing the sensitivity of all considered parameters comprehensively. 

\subsection{Basic characteristics of observation site selection by proposed method}
\label{sec:selection_basic}
In this subsection, the characteristics of observation site selection by the proposed method in the case of the hypocenter~1, and the frequency band of $0.1-1.0$~Hz is investigated in detail as a basic case in the present paper.

\begin{figure*}
    \centering
    \includegraphics[width=13cm]{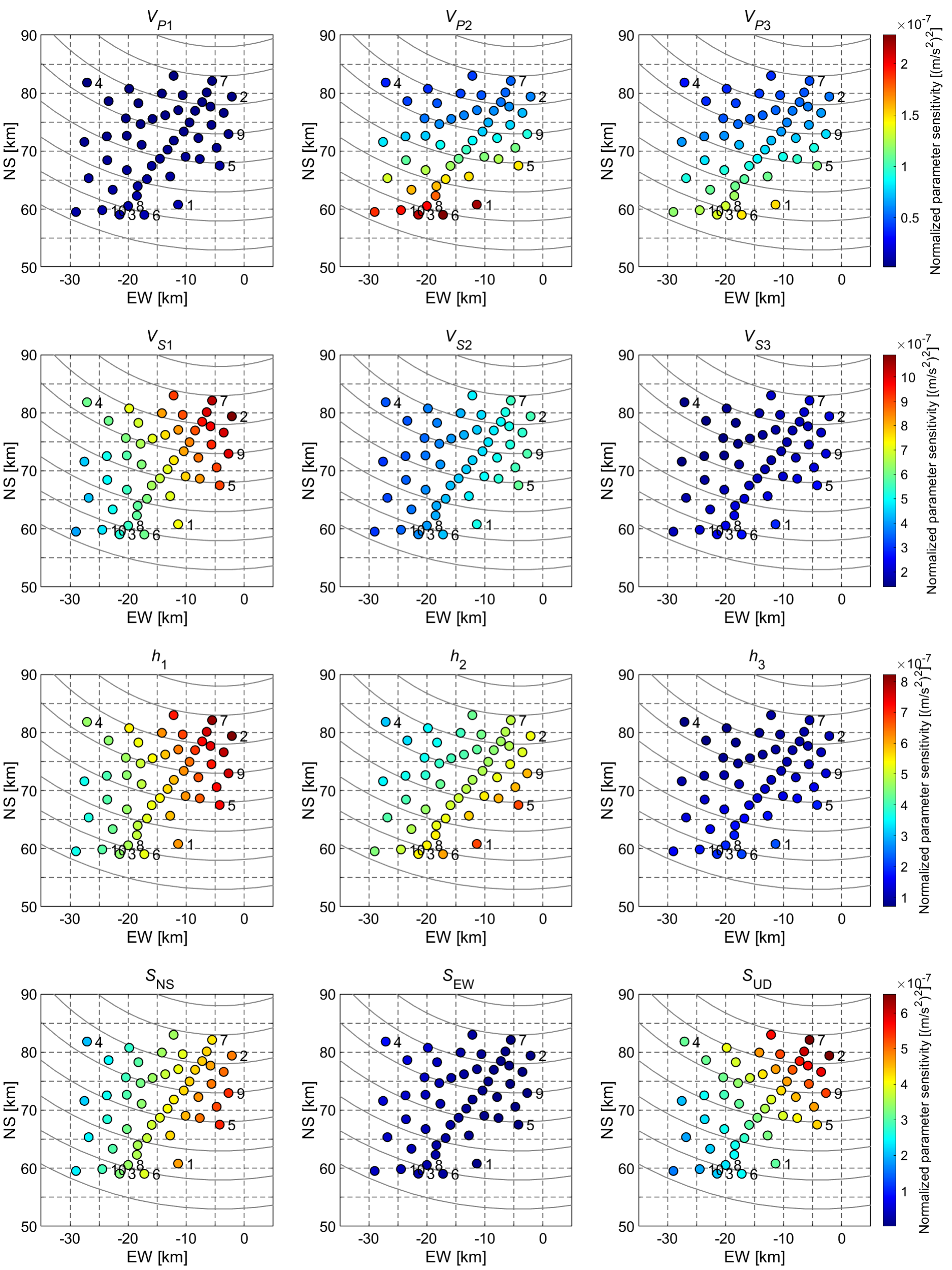}
    \caption{Maps of normalized parameter sensitivity in the case of the hypocenter~1 and frequency band of $0.1-1.0$~Hz. Note that the numbers in the figure indicate the selection order of the top 10 observation sites using the proposed method mentioned in Section~\ref{sec:selection_basic_sel}.}
    \label{fig:case1_sensitivity_map}
\end{figure*}

Figure~\ref{fig:case1_sensitivity_map} shows maps of the normalized parameter sensitivity of 12 parameters. The ranges of color bar in Fig.~\ref{fig:case1_sensitivity_map} were fixed  for each physical parameter. The gray solid lines indicate concentric circles from the hypocenter location every 5~km. Figure~\ref{fig:case1_sensitivity_map} demonstrates that the normalized parameter sensitivity has a spatial structure and there are superiority and inferiority of sensitivity between layers or directions. \color{\ReviewerA}Table \ref{table:factors_hypo1} summarises the characteristics of the parameter sensitivity.\color{black} The characteristics are discussed for each parameter below. 

\begin{table}
\color{\ReviewerA}
    \caption{\color{\ReviewerA}The factors determining the model parameter sensitivity in the case of the hypocentre 1.}
    \label{table:factors_hypo1}
    \begin{tabular}{lll}
    \hline
    Sensitivity & Characteristics & Dominant factor \\
    \hline
    Sensitivity to $V_P$ & \begin{tabular}{l}Magnitude relationship \\between layers\end{tabular} & The phase shift\\
    & \begin{tabular}{l}Spatial structure\end{tabular} & The phase shift\\
    \hline
    Sensitivity to $V_S$ & \begin{tabular}{l}Magnitude relationship \\between layers\end{tabular} & The phase shift\\
    & \begin{tabular}{l}Spatial structure\end{tabular} & The energy distribution of the S wave\\
    \hline
    Sensitivity to $h$ & \begin{tabular}{l}Magnitude relationship between \\layers and spatial structure\end{tabular} & \begin{tabular}{l}Determined by the sensitivity \\to $V_P$ and $V_S$\end{tabular}\\
    \hline
    Sensitivity to $S_{\rm{NS}}$, $S_{\rm{EW}}$, $S_{\rm{UD}}$ & \begin{tabular}{l}Magnitude relationship between \\directions and the spatial structure\end{tabular} & \begin{tabular}{l}Characterized by the positional \\relationship and the energy \\distribution of the S wave\end{tabular}\\
    \hline
    \end{tabular}
\end{table}
\color{black}


\subsubsection{Sensitivity to seismic velocities $V_P$ and $V_S$}
\label{sec:selection_basic_V}
Firstly, the characteristics of the sensitivity to seismic velocities $V_P$ and $V_S$ are discussed. 
Focusing on the magnitude relationship between layers, the sensitivity to $V_{P\rm{2}}$ and $V_{P\rm{3}}$ is much higher than that to $V_{P\rm{1}}$, and the sensitivity to $V_{S\rm{1}}$ and $V_{S\rm{2}}$ is much higher than that to $V_{S\rm{3}}$. The reason for the tendency is considered to be the difference in the arrival time of the seismic wave due to the difference in the seismic velocities. The difference in the seismic velocity of each layer causes the difference in the arrival time, that is, the phase shift, and the greater the difference in arrival time, the higher the sensitivity. Here, the arrival time from the hypocenter location to the observation site on the ground is estimated in the condition utilized in the present study, assuming that the wave goes straight up (in the UD direction) for simplicity, while the wave travels intricately by reflection and refraction. Table~\ref{tab:DiffArrivalTime} shows the difference in the arrival time caused by the difference in the seismic velocity ($2\Delta V_P, 2\Delta V_S$). Here, the values of $\Delta V_P$ and $\Delta V_S$ are set to the same as those for constructing the normalized sensitivity matrix described in Section~\ref{sec:selection_probset}. The magnitude relationship of the differences in the arrival time between 3 layers for $V_P$ and $V_S$ is similar to that of the sensitivity between 3 layers for $V_P$ and $V_S$, respectively.
Therefore, the magnitude relationship of the sensitivity to $V_P$ and $V_S$ of different layers is considered to be characterized by a phase shift due to the difference of $V_P$ and $V_S$.

\begin{table}
    \caption{Estimated result of the difference in the arrival time from the hypocenter location to the observation site on the ground caused by the difference in the seismic velocity ($2\Delta V_P, 2\Delta V_S$). $\Delta V$ is 10\% of the initial values shown in Table~\ref{tab:ModelParam}.}
    \label{tab:DiffArrivalTime}
    \begin{tabular}{lll}
    \hline
     & Diff. in arrival time caused by $2\Delta V_P$ [ms] & Diff. in arrival time caused by $2\Delta V_S$ [ms] \\ \hline
    Layer 1 & 0.0444 & 0.160 \\
    Layer 2 & 0.0916 & 0.220 \\
    Layer 3 & 0.0625 & 0.118 \\ \hline
    \end{tabular}
\end{table}

\begin{figure}
    \centering
    \includegraphics[width=5cm]{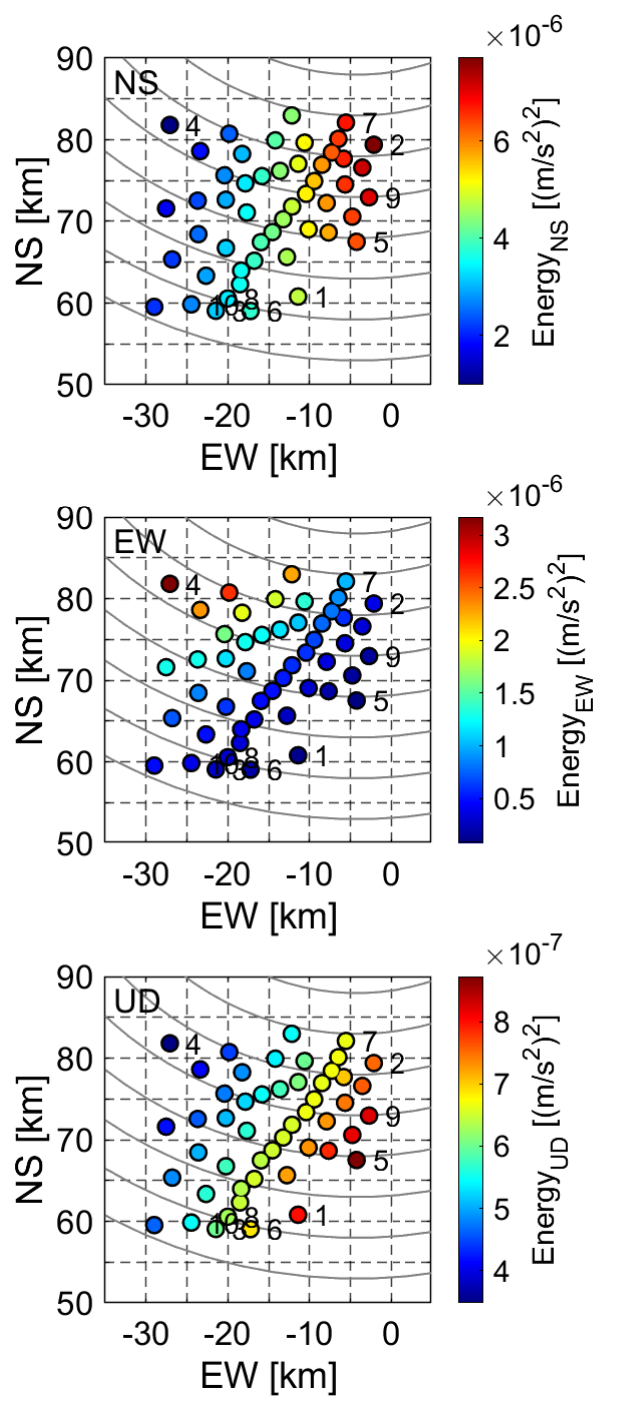}
    \caption{Energy distributions in the NS, EW, and UD directions in the case of the hypocenter~1. Note that the numbers in the figure indicate the selection order of the top 10 observation sites using the proposed site selection method mentioned in Section~\ref{sec:selection_basic_sel}.}
    \label{fig:case1_energy}
\end{figure}

\begin{figure}
    \centering
    \includegraphics[width=4cm]{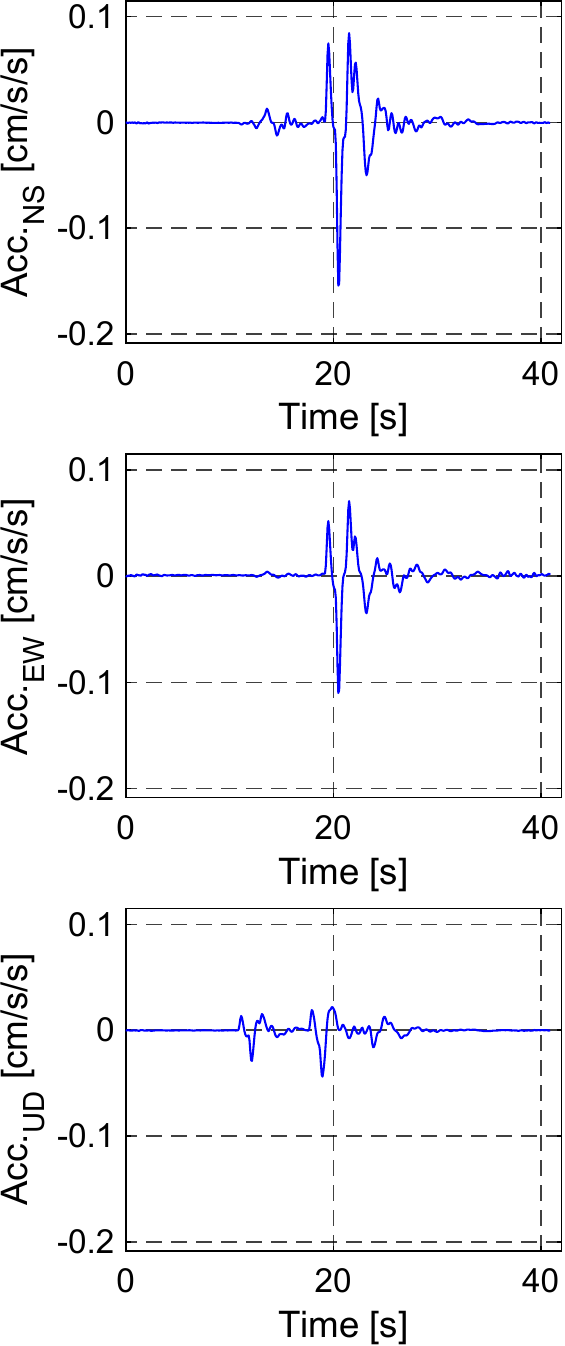}
    \caption{Seismic waveforms in the NS, EW, and UD directions at the observation site with the shortest epicentral distance in the case of the hypocenter~1.}
    \label{fig:case1_waveform}
\end{figure}

Focusing on the spatial structure of the sensitivity to the velocities, the sensitivity to $V_P$ is higher at observation sites with a larger epicentral distance. This trend is due to the phase shift similar to the magnitude relationship; the difference in the arrival time increases with increasing the epicenter distance of observation site. 
On the other hand, the sensitivity to $V_S$ is higher at the observation sites closer to the northeast. This trend is due to the energy distribution of the S-wave. Figure~\ref{fig:case1_energy} shows the energy distributions in the NS, EW, and UD directions. The energy in each direction is obtained by calculating the sum of squares of the components in each direction of frequency spectra. Figure~\ref{fig:case1_waveform} shows the waveforms in the NS, EW, and UD directions sampled at the observation site with the shortest epicentral distance as an example. Figure~\ref{fig:case1_waveform} illustrates that the S wave is dominant compared with the P-wave, and the fluctuation in the NS direction is dominant in the duration of S-wave arrival in the current case. Thus, the energy distribution in the NS direction (Fig.~\ref{fig:case1_energy}) corresponds to that of S-wave. 
The comparison of the energy distribution in the NS direction (Fig.~\ref{fig:case1_energy}) with the maps of sensitivity to $V_S$ (Fig.~\ref{fig:case1_sensitivity_map}) demonstrates that the spatial structure of the sensitivity to $V_S$ is qualitatively the same as that of the energy distribution of the S-wave. Therefore, the reason for the trend in the spatial structure is that the difference in the energy of the S-wave caused by the difference in $V_S$ increases with originally increasing the energy of the S-wave.
As discussed above, the spatial structure of the sensitivity to $V_P$ and $V_S$ is characterized by the phase shift due to the difference in the velocities and the energy distribution of the S-wave. The effects of the phase shift and the energy distribution of the S-wave are dominant for $V_P$ and $V_S$, respectively, in the case of the hypocenter~1.

\subsubsection{Sensitivity to layer thickness $h$}
\label{sec:selection_basic_h}
Secondly, the characteristics of the sensitivity to layer thickness $h$ are discussed. 
With regard to the magnitude relationship, the sensitivity to $h_{\rm{1}}$ and $h_{\rm{2}}$ is much higher than that to $h_{\rm{3}}$. Focusing on the spatial structure, the sensitivity to $h_{\rm{1}}$ is higher at observation sites closer to the northeast, and that to $h_{\rm{2}}$ is higher at observation sites closer to the southeast. 
These characteristics of the sensitivity to $h$ are considered to be strongly related to the sensitivity to seismic velocities $V_P$ and $V_S$ because the difference in the layer thickness has a qualitatively similar effect on the arrival time to that in the seismic velocities. The spatial structure of the sensitivity to $h_{\rm{1}}$ is qualitatively the same as that to $V_{S\rm{1}}$, which is relatively high while the sensitivity to $V_{P\rm{1}}$ is low. 
With regard to the layer~2, the sensitivity to $V_{P\rm{2}}$ and $V_{S\rm{2}}$ is both relatively high; thus, the characteristics of the sensitivity to $V_{P\rm{2}}$ and $V_{S\rm{2}}$ are superimposed in the spatial structure of the sensitivity to $h_{\rm{2}}$. 
On the other hand, With regard to the layer~3, the sensitivity to $h_{\rm{3}}$ is low since both sensitivity to $V_{P\rm{3}}$ and $V_{S\rm{3}}$ are relatively low. 
Therefore, the characteristics of the sensitivity to $h$ are determined by those to $V_P$ and $V_S$.

\subsubsection{Sensitivity to hypocenter locations $S_{\mathrm{NS}}$, $S_{\mathrm{EW}}$, and $S_{\mathrm{UD}}$}
 \label{sec:selection_basic_S}
Finally, the characteristics of the sensitivity to the hypocenter locations $S_{\mathrm{NS}}$, $S_{\mathrm{EW}}$, and $S_{\mathrm{UD}}$ are discussed. 
With regard to the magnitude relationship, the sensitivity to $S_{\mathrm{NS}}$ and $S_{\mathrm{UD}}$ is much higher than that of $S_{\mathrm{EW}}$. Focusing on the spatial structure, the sensitivity to $S_{\mathrm{NS}}$ is higher at observation sites closer to the eastern, and that to $S_{\mathrm{UD}}$ is higher at observation sites closer to the northeastern. 
These characteristics are due to the positional relationship and the energy distribution of the S-wave. The hypocenter is far north of the area of observation sites. In addition, the fluctuation in the NS direction is confirmed to be dominant in the duration of S-wave arrival as described in Section~\ref{sec:selection_basic_V}. 
Consequently, when the hypocenter location shifts in the NS direction, the observation sites closer to the same EW coordinates as the hypocenter are more suitable for observing the change in the fluctuation in the NS direction. This is the reason for the spatial structure of the sensitivity to $S_{\mathrm{NS}}$.
On the other hand, when the hypocenter location shifts in the EW direction, the effect on the fluctuation at the observation sites, especially the fluctuation in the NS direction, is relatively small. This is the reason why the sensitivity to $S_{\mathrm{EW}}$ is low.
Furthermore, the spatial structure of the sensitivity to $S_{\mathrm{UD}}$ is qualitatively the same as that of the energy distribution of the S-wave shown in Fig.~\ref{fig:case1_energy}. This is because the larger the energy of the S-wave, the larger the energy difference due to the difference in model parameters, as discussed in Section~\ref{sec:selection_basic_V}. Note that although the sensitivity to $S_{\mathrm{NS}}$ and $S_{\mathrm{EW}}$ is also influenced by the energy difference, the dominant factor is considered to be the positional relationship rather than the energy difference.
Therefore, the characteristics of the sensitivity to the hypocenter locations are characterized by the positional relationship and the energy distribution of the S-wave. The effect of the positional relationship is dominant for $S_{\mathrm{NS}}$ and $S_{\mathrm{EW}}$, and that of the energy distribution of the S-wave is dominant for $S_{\mathrm{UD}}$, in the case of the hypocenter~1.

\subsubsection{Results of observation site selection}
\label{sec:selection_basic_sel}
\begin{figure}
    \centering
    \includegraphics[width=5.5cm]{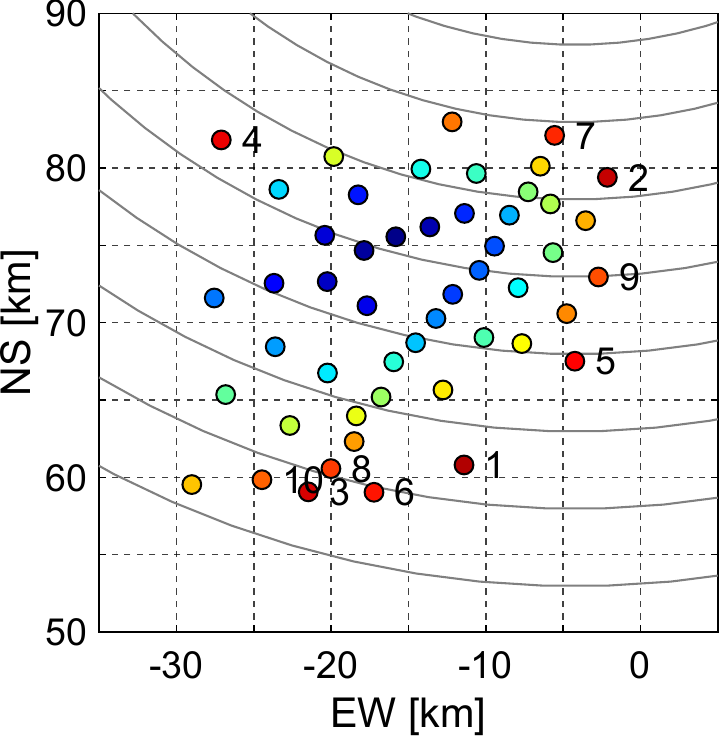}
    \caption{Selection order of observation sites using the proposed method in the case of the hypocenter~1 and frequency band of $0.1-1.0$~Hz. The observation sites are colored from red to blue depending on the order of selection; the observation sites selected earlier are plotted in red, and those selected later are plotted in blue. The numbers in the figure indicate the selection order of the top 10 observation sites.}
    \label{fig:case1_obsposi}
\end{figure}

\begin{figure*}
    \centering
    \includegraphics[width=13cm]{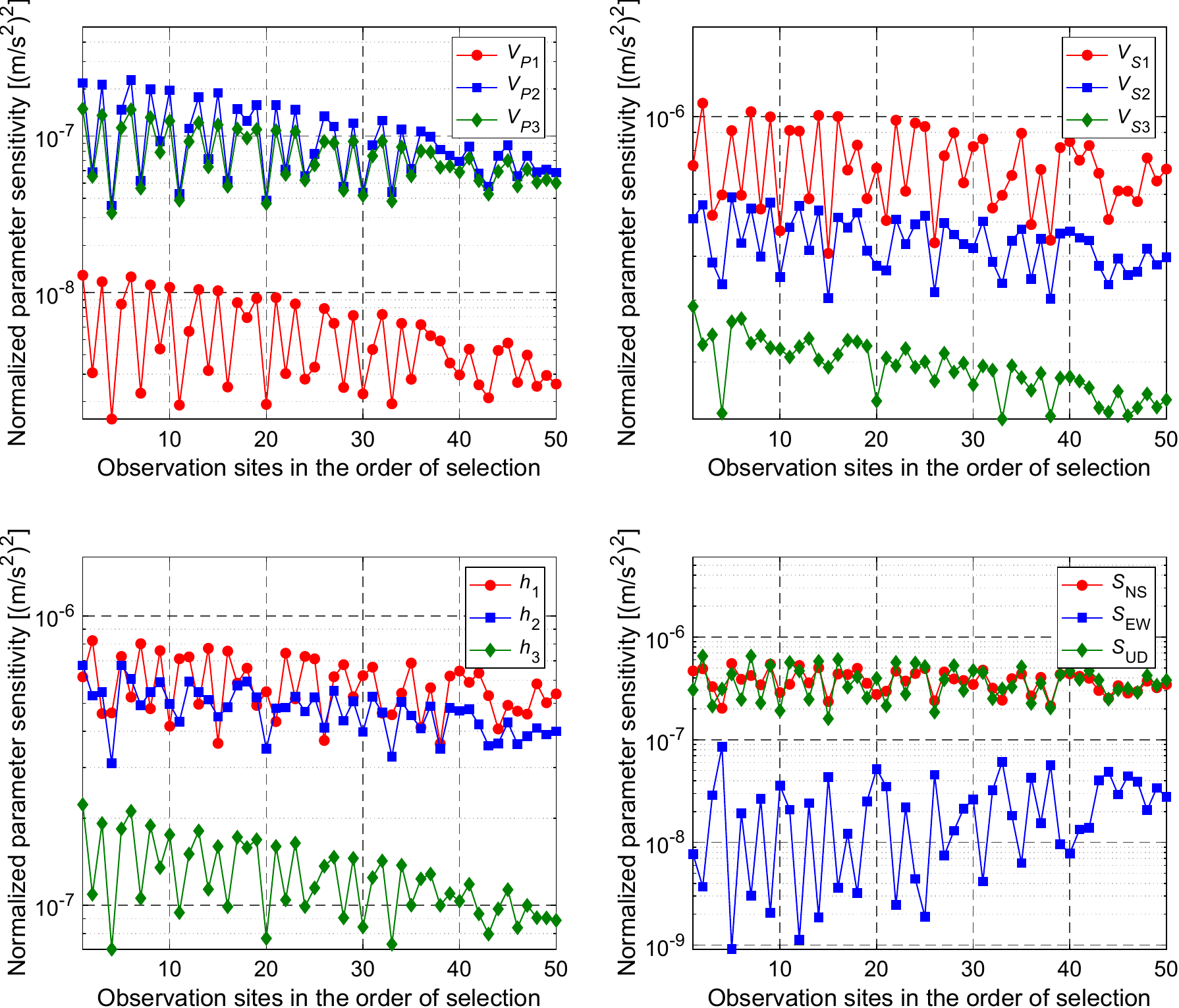}
    \caption{Normalized parameter sensitivity against observation sites in the order of selection in the case of the hypocenter~1 and frequency band of $0.1-1.0$~Hz.}
    \label{fig:case1_sensitivity_semilog}
\end{figure*}

Next, the results of observation site selection are discussed. 
Figure~\ref{fig:case1_obsposi} shows the result of observation site selection using the proposed method. The top 10 observation sites are numbered in the order of selection in Fig.~\ref{fig:case1_obsposi}. Moreover, the top 10 observation sites are numbered in Figs.~\ref{fig:case1_sensitivity_map} and \ref{fig:case1_energy} to show the relevance to the normalized parameter sensitivity. Figure~\ref{fig:case1_sensitivity_semilog} shows the normalized parameter sensitivity against observation sites in the order of selection. 
Figure~\ref{fig:case1_obsposi} indicates that observation sites outside the area of observation sites tend to be preferentially selected by the proposed method. This trend is due to the characteristics of the normalized parameter sensitivity discussed above; Fig.~\ref{fig:case1_sensitivity_semilog} demonstrates that the values of sensitivity decrease comprehensively as the selection order of observation sites decreases. These results confirm that the proposed method preferentially selects observation sites with high sensitivity to the parameters. 
On the other hand, some parameters have different tendencies (e.g., $S_{\mathrm{EW}}$). This is because the proposed method emphasizes the D-optimality criterion of the sensitivity matrix, which represents the average sensitivity of all the parameters, rather than directly evaluating the normalized parameter sensitivity, which corresponds to the sensitivity of each parameter. Consequently, when there is a parameter whose spatial structure is qualitatively different from the major structure of all parameters, the order of observation site selection may not correlate with the sensitivity to the parameter.

\subsection{Effect of frequency band}
\label{sec:selection_freq}
\begin{figure*}
    \centering
    \includegraphics[width=13cm]{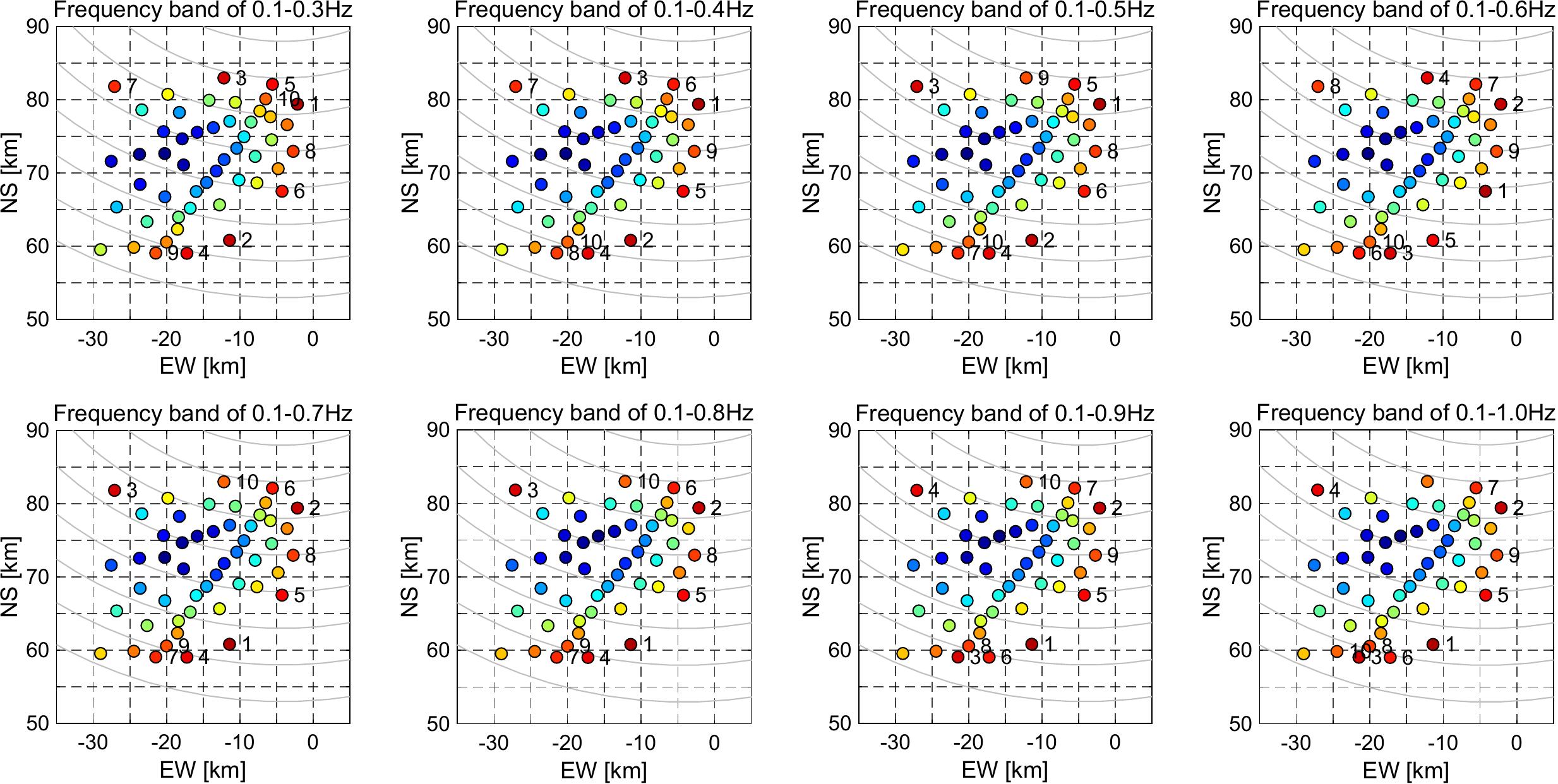}
    \caption{Comparison of the order of observation site selection using the proposed method in the cases of different frequency bands.}
    \label{fig:case1_freqcomp_obsposi}
\end{figure*}

\begin{figure*}
    \centering
    \includegraphics[width=13cm]{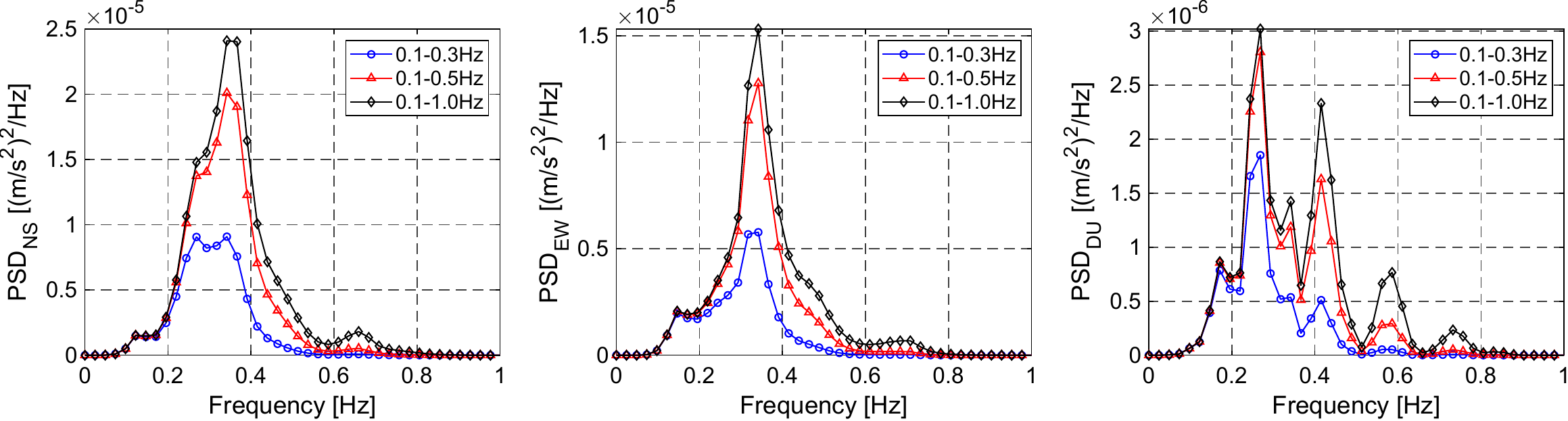}
    \caption{Comparison of the power spectral density in the cases of different frequency bands.}
    \label{fig:case1_freqcomp_psd}
\end{figure*}

In this subsection, the effect of the frequency band of the seismic wavefield on the observation site selection is investigated.
Figure~\ref{fig:case1_freqcomp_obsposi} shows the results of observation site selection using the proposed method in the cases of different frequency bands. The higher cutoff frequency of the bandpass filter varies from 0.3 to 1.0~Hz in each figure plotted at from the top left to the bottom right in Fig.~\ref{fig:case1_freqcomp_obsposi}. Note that the result of the higher cutoff frequency of 1.0~Hz is the same as the result shown in Fig.~\ref{fig:case1_obsposi}. 
The comparison of the power spectral density of acceleration in the NS, EW, and UD directions is shown in Fig.~\ref{fig:case1_freqcomp_psd}. The power spectral density is sampled at the observation site with the shortest epicentral distance as an example. 
The results using the bandpass filter with the higher cutoff frequency of 0.3, 0.5, and 1.0~Hz are plotted together for comparison in Fig.~\ref{fig:case1_freqcomp_psd}. 
The comparison of the observation site selection in the cases of different frequency bands shown in Fig.~\ref{fig:case1_obsposi} demonstrates that there is no effect of the frequency band on the observation site selection. Although the order of selection is slightly different depending on the frequency band, the comprehensive characteristics discussed in Section~\ref{sec:selection_basic} is not affected by the frequency band. 
This is because the comprehensive characteristics of the seismic wavefield are the same between the cases of the different frequency bands. Figure~\ref{fig:case1_freqcomp_psd} illustrates that the frequency spectra are quantitatively different depending on the frequency band; the values of the power spectral density at the dominant frequency increase significantly with increasing the higher cutoff frequency. However, the qualitative characteristics in the cases of 0.3, 0.5, and 1.0~Hz are the same as each other; the power spectral density is much higher in the NS direction than in the EW and UD directions, and the peak frequencies of the acceleration in the NS, EW, and UD directions does not change.  
In addition, maps of the normalized parameter sensitivity are confirmed not to be qualitatively affected by the frequency band.
Therefore, the observation sites sensitive to the parameters are not considered to depend on the frequency band of the seismic wavefield. 

\subsection{Effect of focal mechanism}
\label{sec:selection_mechanism}
In this subsection, the effect of the focal mechanisms on the observation site selection is discussed. 
The results in the case of the hypocenter~2 are compared to those in the case of the hypocenter~1 discussed in Section~\ref{sec:selection_basic}. 

\begin{figure*}
    \centering
    \includegraphics[width=13cm]{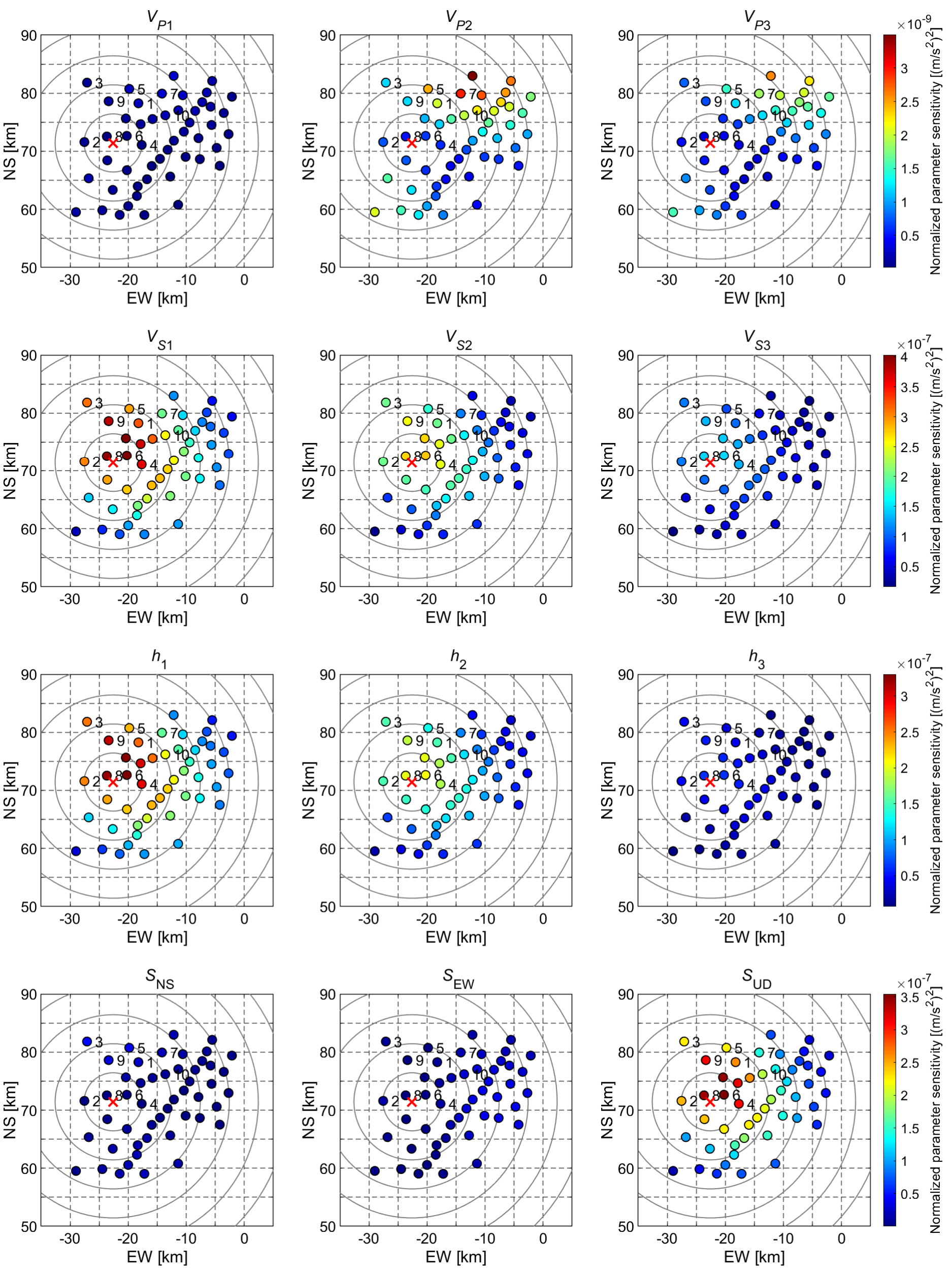}
    \caption{Maps of normalized parameter sensitivity in the case of the hypocenter~2 for frequency band of 0.1--1.0~Hz. Note that the numbers in the figure indicate the selection order of the top 10 observation sites using the proposed method.}
    \label{fig:case2_sensitivity_map}
\end{figure*}

Figure~\ref{fig:case2_sensitivity_map} shows maps of the normalized parameter sensitivity of 12 parameters. The range of the color bar in Fig.~\ref{fig:case2_sensitivity_map} is set to be the same as each other for the same kind of physical parameters of different layers or directions. The gray lines indicate concentric circles from the hypocenter location.
\color{\ReviewerA}Table \ref{table:factors_hypo2} summarises the effects of the focal mechanisms, and each parameter in the table is explained below.\color{black}

\begin{table}
\color{\ReviewerA}
    \caption{\color{\ReviewerA}The factors determining the model parameter sensitivity in the case of the hypocentre 2.}
    \label{table:factors_hypo2}
    \begin{tabular}{llll}
        \hline
        Sensitivity & Characteristics & Dominant factor & Effect of focal mechanisms \\
        \hline
        $V_P$ & \begin{tabular}{l}Magnitude relationship \\between layers\end{tabular} & \begin{tabular}{l}The phase shift\end{tabular} & No \\
        & \begin{tabular}{l}Spatial structure\end{tabular} & \begin{tabular}{l}The energy distribution \\of the P-wave\end{tabular} & \begin{tabular}{l}Yes (Cf. the phase \\shift in the case of the \\hypocenter 1)\end{tabular}\\
        \hline
        $V_S$ & \begin{tabular}{l}Magnitude relationship \\between layers\end{tabular} & \begin{tabular}{l}The phase shift\end{tabular} & No \\
        & \begin{tabular}{l}Spatial structure\end{tabular} & \begin{tabular}{l}The energy distribution \\of the S-wave\end{tabular} & No\\
        \hline
        $h$ & \begin{tabular}{l}Magnitude relationship \\between layers and \\spatial structure\end{tabular} & \begin{tabular}{l}Determined by the \\sensitivity to $V_P$ and $V_S$\end{tabular} & No\\
        \hline
        $S_{\rm{NS}}$, $S_{\rm{EW}}$, $S_{\rm{UD}}$ & \begin{tabular}{l}Magnitude relationship \\between directions and \\the spatial structure\end{tabular} & \begin{tabular}{l}Characterized by the \\positional relationship \\and the energy \\distribution of the S-wave\end{tabular} & No \\
        \hline
    \end{tabular}
\end{table}
\color{black}

\begin{figure}
    \centering
    \includegraphics[width=5.5cm]{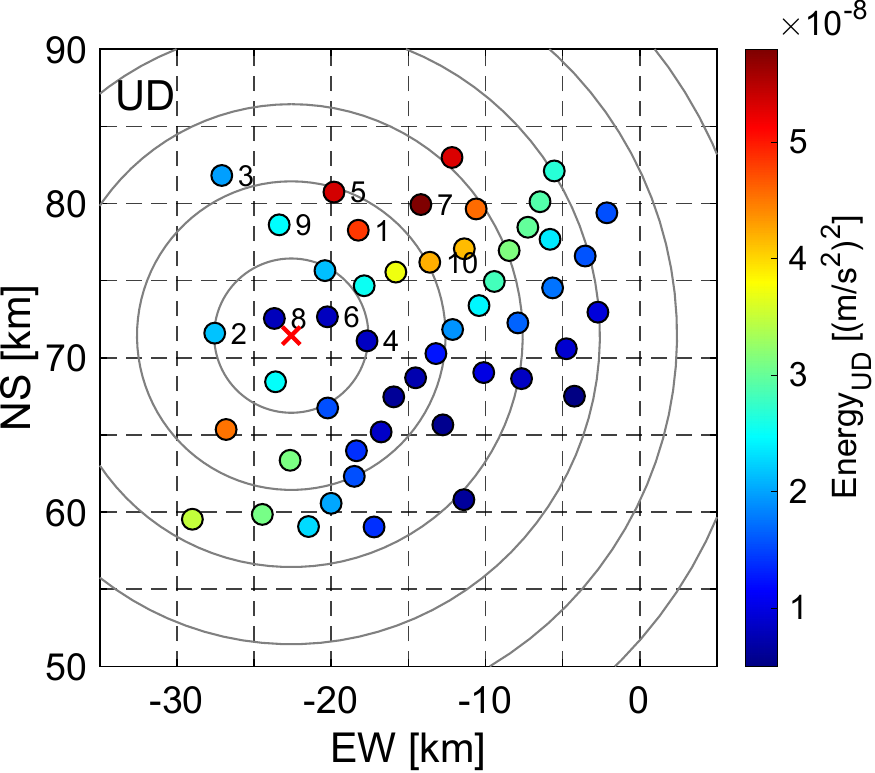}
    \caption{Energy distribution in the UD direction in the case of the hypocenter~2. Note that the numbers in the figure indicate the selection order of the top 10 observation sites using the proposed method.}
    \label{fig:case2_DUenergy}
\end{figure}

Firstly, the characteristics of the sensitivity to seismic velocities $V_P$ and $V_S$ are discussed.
Focusing on the magnitude relationship between layers, the sensitivity to $V_{P\rm{2}}$ and $V_{P\rm{3}}$ are much higher than that to $V_{P\rm{1}}$, and the sensitivity to $V_{S\rm{1}}$ and $V_{S\rm{2}}$ are much higher than that to $V_{S\rm{3}}$. These tendencies are similar to those of the case of the hypocenter~1 in Section~\ref{sec:selection_basic_V}; the magnitude relationship is due to the difference in the arrival time. 
Therefore, the magnitude relationship of the sensitivity to seismic velocities between layers does not depend on the focal mechanisms and is determined by the subsurface structure model parameters. 
Focusing on the spatial structure, the sensitivity to $V_S$ is high in the northern region adjacent to the hypocenter. The reason for the characteristics is consistent with the discussion on the case of the hypocenter~1 in Section~\ref{sec:selection_basic_V}; the spatial structure of the sensitivity to $V_S$ is due to the energy distribution of the S-wave. 
On the other hand, the characteristics of the spatial structure of the sensitivity to $V_P$ are different; although the sensitivity to $V_P$ is higher at observation sites with a larger epicentral distance in the case of the hypocenter~1, it is high in the north region and the southwest region in the case of the hypocenter~2. This is due to the energy distribution of the P-wave. Figure~\ref{fig:case2_DUenergy} shows the energy distribution in the UD direction obtained by calculating the sum of squares of the components in the UD direction of the frequency spectra. The energy distribution corresponds to that of the P-wave because the fluctuations in the NS and EW directions are dominant in the duration of the S-wave arrival and that in the UD direction is dominant in the duration of the P-wave arrival in the current case. The comparison of the energy distribution (Fig.~\ref{fig:case2_DUenergy}) with the maps of sensitivity to $V_P$ (Fig.~\ref{fig:case2_sensitivity_map}) demonstrates that the spatial structure of the sensitivity to $V_P$ is qualitatively the same as that of the energy distribution of the P-wave. Hence, the reason for the trend of the spatial structure is that the difference in the energy of the P-wave caused by the difference in $V_P$ increases with increasing originally the energy of the P-wave. 

To summarize the discussions on the spatial structure of the sensitivity to $V_P$ above and in Section~\ref{sec:selection_basic_V}, the difference in the arrival time and the energy distribution of the P-wave are the dominant factors in the cases of the hypocenters~1 and 2, respectively. The discrepancy is mainly because the hypocenter is located approximately 30 to 50 km away from the observation site area in the case of the hypocenter~1, whereas in the case of the hypocenter~2, the hypocenter is directly below the observation site area, and the difference in the arrival time between the observation sites is relatively small. 
Therefore, factors that potentially affect the spatial structure of the sensitivity to seismic velocities do not depend on the focal mechanism; the difference in the arrival time and the energy distribution affects the sensitivity in both cases of hypocenter~1 and 2, but the dominant factor depends on the focal mechanism.

Secondly, the characteristics of the sensitivity to $h$ are discussed. 
The sensitivity to $h_{\rm1}$ and $h_{\rm2}$ is higher than that to $h_{\rm3}$, and it is high in the northern region adjacent to the hypocenter. The reasons for these characteristics are consistent with the discussions on the case of the hypocenter~1 in Section~\ref{sec:selection_basic_h}; the characteristics of the sensitivity to $h$ are determined by those to $V_P$ and $V_S$. 
Therefore, factors that affect the characteristics of the sensitivity to $h$ do not depend on the focal mechanisms.

Finally, the characteristics of the sensitivity to $S_{\mathrm{NS}}$, $S_{\mathrm{EW}}$, and $S_{\mathrm{UD}}$ are discussed.
The sensitivity to $S_{\mathrm{UD}}$ is much higher than that of $S_{\mathrm{NS}}$ and $S_{\mathrm{EW}}$, and $S_{\mathrm{UD}}$ is higher in the northern region adjacent to the hypocenter. The reasons for these characteristics are consistent with the discussions on the case of the hypocenter~1 in Section~\ref{sec:selection_basic_S}; the characteristics of the sensitivity to the hypocenter locations are characterized by the positional relationship and the energy distribution of the S-wave. 
Therefore, factors that potentially affect the characteristics of the sensitivity to hypocenter locations do not depend on the focal mechanism. 

\begin{figure}
    \centering
    \includegraphics[width=5.5cm]{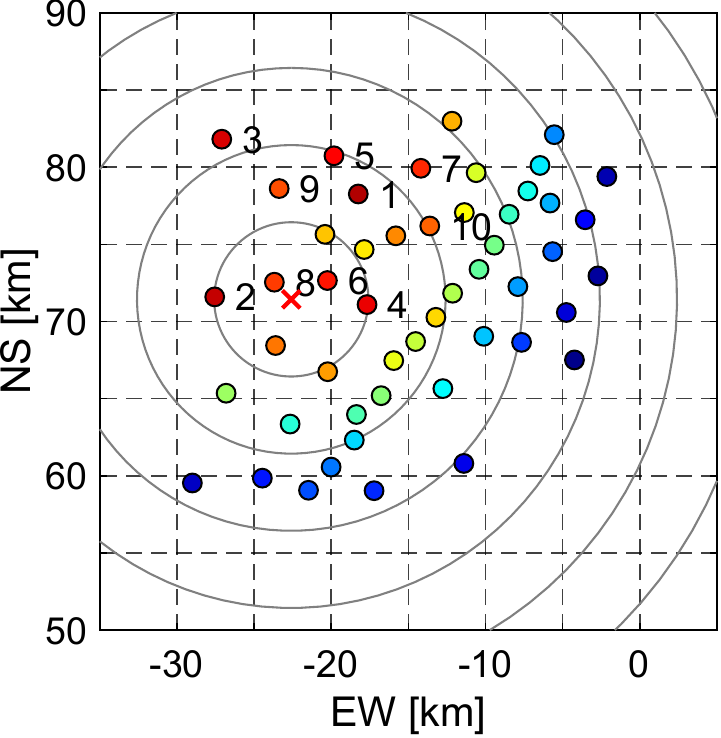}
    \caption{Selection order of observation sites using the proposed method in the case of the hypocenter~2 and frequency band of 0.1--1.0~Hz. The observation sites are colored from red to blue depending on the order of selection. The numbers in the figure indicate the selection order of the top 10 observation sites.}
    \label{fig:case2_obsposi}
\end{figure}

\begin{figure*}
    \centering
    \includegraphics[width=13cm]{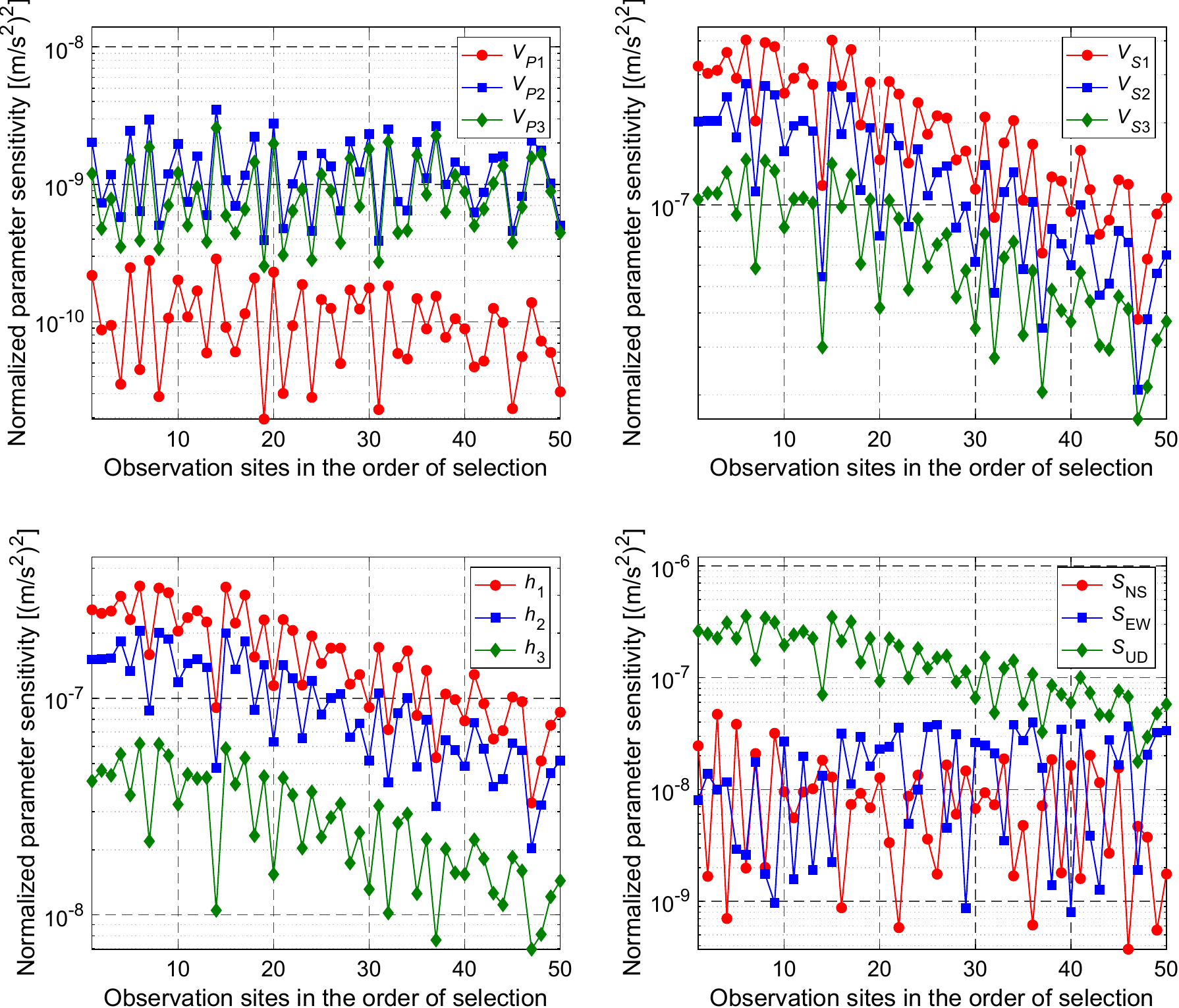}
    \caption{Normalized parameter sensitivity against observation sites in the order of selection in the case of the hypocenter~2 and frequency band of 0.1--1.0~Hz.}
    \label{fig:case2_sensitivity_semilog}
\end{figure*}

Figure~\ref{fig:case2_obsposi} shows the result of observation site selection using the proposed method in the case of the hypocenter~2 and the frequency band of $0.1$--1.0~Hz. The observation sites are colored from red to blue depending on the order of selection. Figure~\ref{fig:case2_sensitivity_semilog} shows the normalized parameter sensitivity against observation sites in the order of selection.
Figure~\ref{fig:case2_obsposi} indicates that observation sites in the northern region adjacent to the hypocenter are preferentially selected by the proposed method. This trend is due to the characteristics of the normalized parameter sensitivity discussed above; Fig.~\ref{fig:case2_sensitivity_semilog} demonstrates that the values of sensitivity decrease comprehensively as the selection order of observation sites decreases. 
Especially in the case of the hypocenter~2, the sensitivity to $V_S$, $h$, and $S_{\mathrm{UD}}$ decreases with the selection order, whereas there is no correlation between the selection order and other parameters such as $V_P$, $S_{\mathrm{NS}}$, and $S_{\mathrm{EW}}$. This is mainly because the spatial structure of the sensitivity to $V_S$, $h$, and $S_{\mathrm{UD}}$ is major among all parameters, whereas those of other parameters are different from the major one. 
Therefore, the proposed method is confirmed to be able to select observation sites suitable for improvement of the average sensitivity to all parameters according to the sensitivity characteristics, which depends on the focal mechanism. 

\section{Twin experiment of seismic wavefield reconstruction}
\label{sec:twinexp}
A numerical experiment of the seismic wavefield reconstruction is conducted using synthetic observation data, and the proposed method is verified. The synthetic observation data are theoretical seismic waveforms calculated using the code developed by \citet{hisada1995efficient} described in Section~\ref{sec:methodologies_selection}. Note that the code developed by \citet{hisada1995efficient} is used not only for selecting the observation site and reconstructing the seismic wavefield but also for preparing true data for the seismic wavefield reconstruction; therefore, a twin experiment, where both true and estimated waveforms are calculated using the same simulator, is conducted in the present study. 
First, the observation sites are selected using the proposed method based on the observation data calculated using the initial values of the model parameters, similar to Section~\ref{sec:selection}. 
Then, the model parameters estimation is conducted based on the observed data at selected observation sites in the way described in Section~\ref{sec:methodologies_reconst}, whereas a slight difference between the initial and true values of the model parameters is assumed.  
Finally, the seismic wavefield is reconstructed using the estimated model parameters.
We verify the accuracy of the model parameter estimation using the observation sites selected by the proposed method by comparing it with that using the randomly selected observation sites. Furthermore, the accuracy of the seismic wavefield reconstruction with the estimated model parameters using the observation sites selected by the proposed method is compared with that using the randomly selected observation sites.

\subsection{Problem settings for twin experiment} \label{sec:twinexp_probset}
\begin{figure*}
    \centering
    \includegraphics[width=13cm]{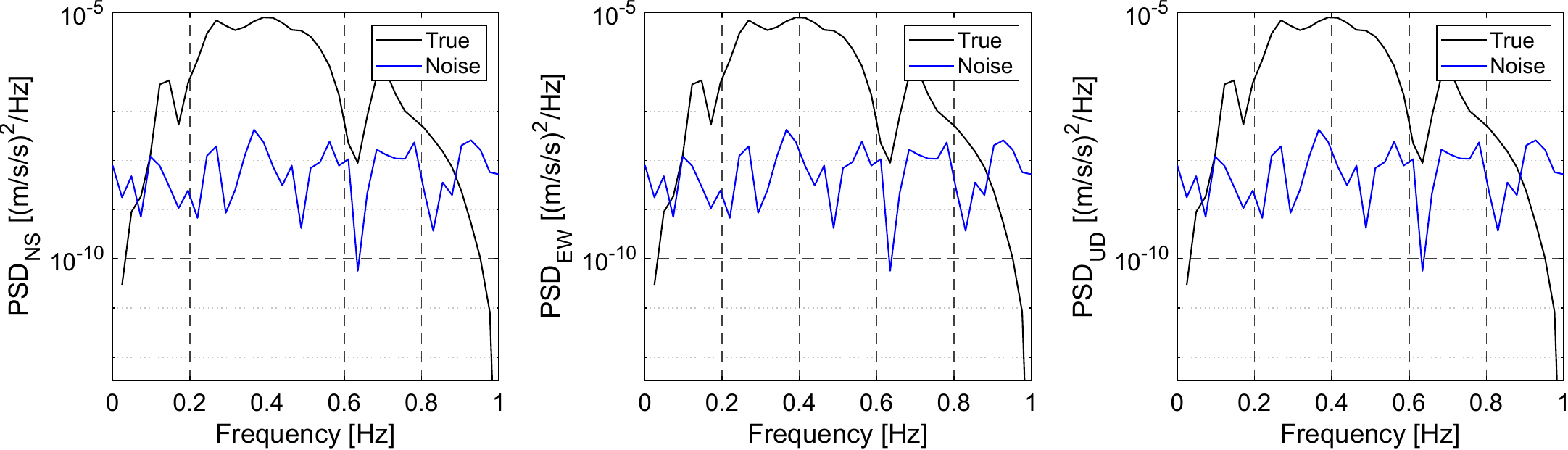}
    \caption{Comparison of the power spectral density of the true and additive noise signal at the observation site with the shortest epicentral distance.}
    \label{fig:twinexp_psd_noise}
\end{figure*}

The seismic velocities ($V_p$ and $V_s$) and the thickness ($h$) of each of the three layers, and source location in three directions ($S_{\mathrm{NS}}$, $S_{\mathrm{EW}}$, and $S_{\mathrm{UD}}$) are employed as the target model parameters to be estimated as with Section~\ref{sec:selection}.
The observation sites are the same as that in Section~\ref{sec:selection_probset}. Hence, the initial values of the subsurface structure model parameters are based on \citet{koketsu2011unified} summarized in Table~\ref{tab:ModelParam}. The earthquake source information is the same as the case of the hypocenter~1 in Section~\ref{sec:selection}, and the initial values of the source parameters are summarized in Table~\ref{tab:HypoParam}.
True values of the 12 model parameters are assumed to be slightly different from the initial values in the numerical experiment. The true values are obtained by adding the small values to the initial values. The small value is obtained by multiplying the random value following a normal distribution $\mathcal{N}(0,1)$ by $10\%$ of the initial value and 5~km for subsurface structure model parameters and source location parameters, respectively. 
The reconstruction of waveforms within a frequency band of 0.1--1.0~Hz is examined in the present experiment. Thus, the simulation was conducted in the frequency range less 5~Hz as with Section~\ref{sec:selection_probset}, and the second-order Butterworth filter of the frequency band of 0.1--1.0~Hz was applied to the obtained simulation results. The observation noise is added to the observation data $\mathbf{y}_p$ calculated using the code developed by \citet{hisada1995efficient} in the frequency domain; the amplitude of additive noise for each component of observed data (each entry of $\mathbf{y}_p$) follows a normal distribution $\mathcal{N}(0,10^{-5})$. Figure~\ref{fig:twinexp_psd_noise} shows the power spectral density in the NS, EW, and UD directions of the true and additive noise signals. The true signals are obtained by the simulation using the true model parameters. 
Three observation sites out of  50 candidates are used for the parameter estimation. Hence, the observation sites numbered 1, 2, and 3 in Fig.~\ref{fig:case1_obsposi} are employed in the proposed method. The experiments using 100 patterns of subsets of three randomly selected observation sites are conducted for comparison. Note that the results of the comparison between the proposed method and averaged results with random selection in the case of three observation sites are qualitatively the same regardless of the number of observation sites, and the superiority of the proposed method tends to increase as the number of observation sites decreases. 

\subsection{Results of twin experiment} \label{sec:twinexp_result}
\begin{figure}
    \centering
    \includegraphics[width=5.5cm]{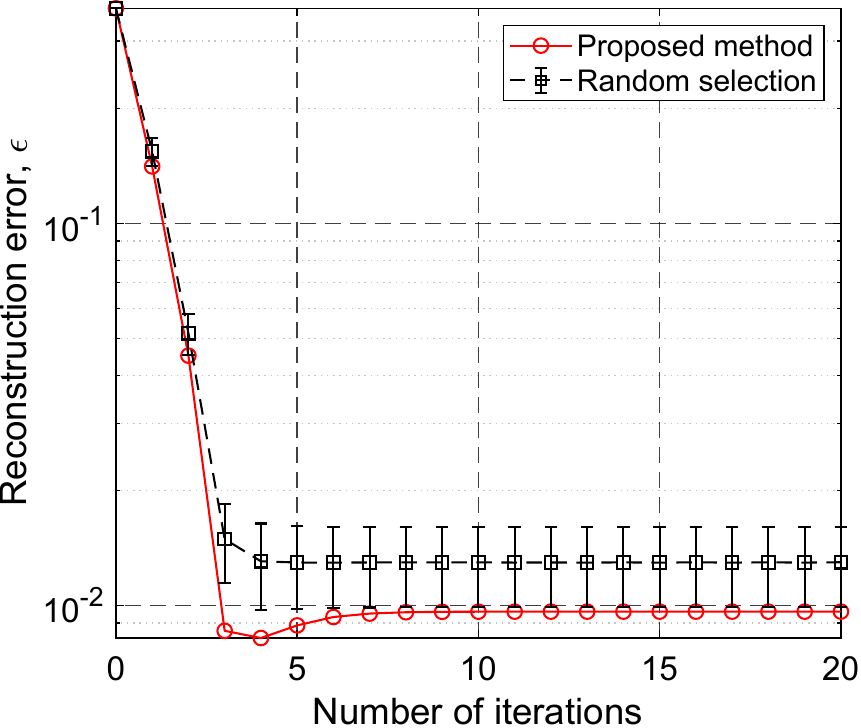}
    \caption{Comparison of the reconstruction error using the three observation sites selected by the proposed method and using three randomly selected observation sites.}
    \label{fig:twinexp_err}
\end{figure}

\begin{figure}
    \centering
    \includegraphics[width=13cm]{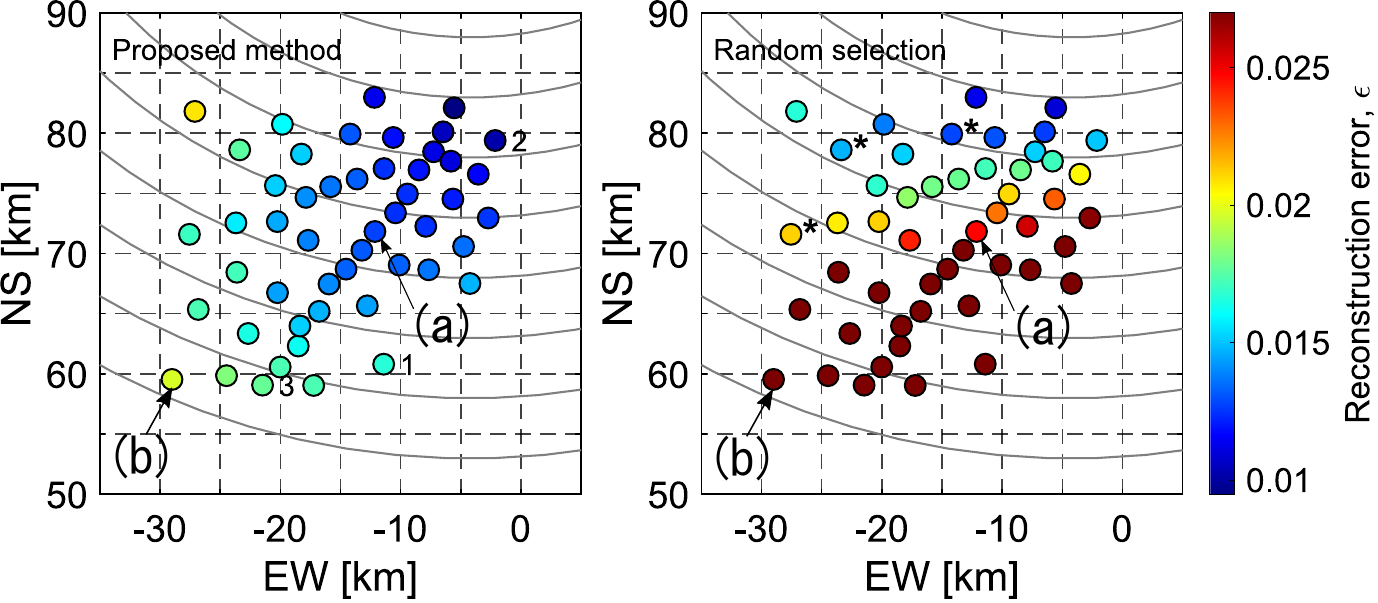}
    \caption{\color{\ReviewerA}The reconstruction error distributions with the model parameters estimated (left) using the three observation sites selected by the proposed method and (right) using three randomly selected observation sites.\color{black}}
    \label{fig:twinexp_errmap}
\end{figure}

\begin{figure*}
    \centering
    \includegraphics[width=13cm]{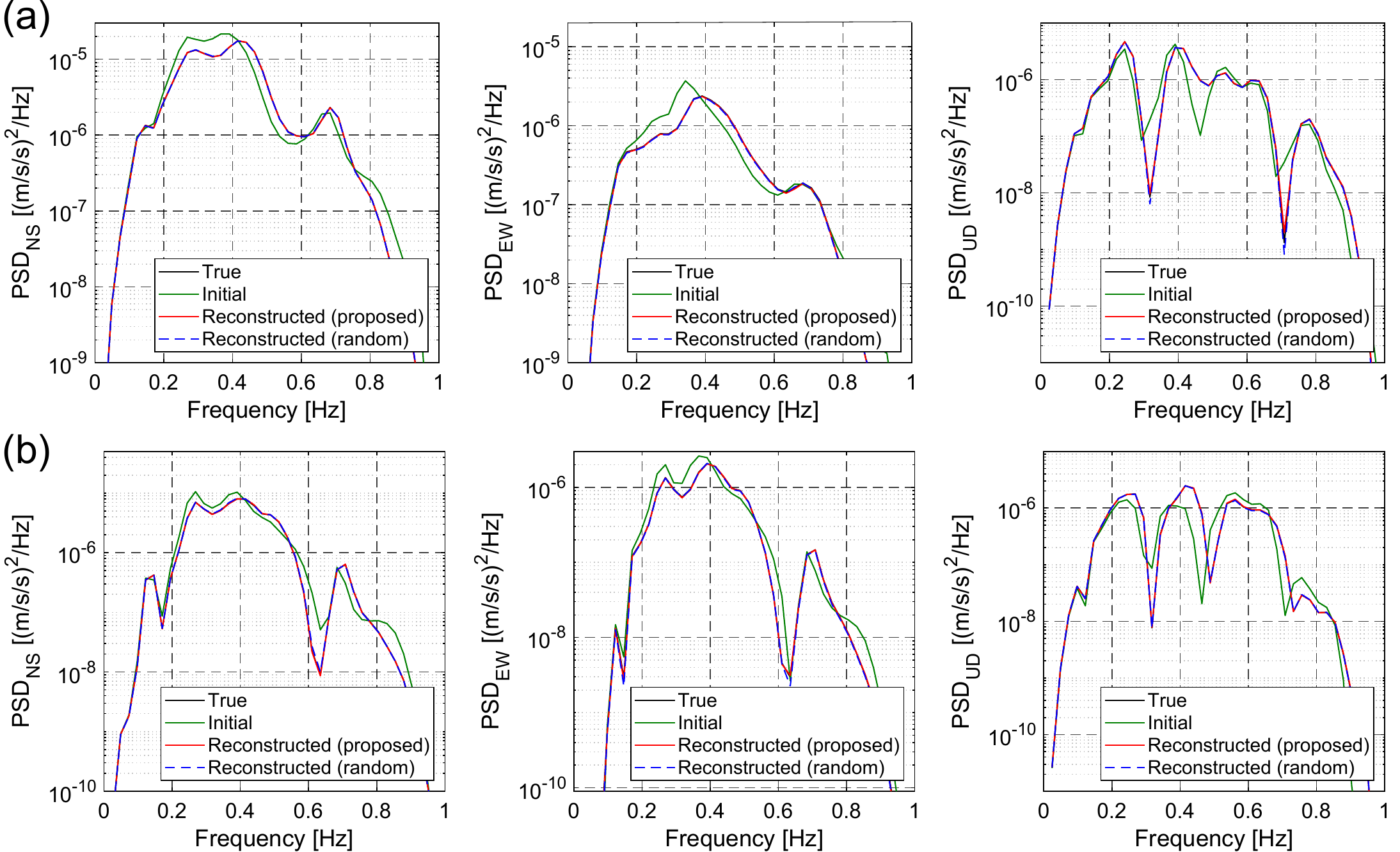}
    \caption{\color{\ReviewerA}Comparison of the power spectral density reconstructed with the true parameters, reconstructed with the initial parameters, reconstructed with the parameters estimated using the three observation sites selected by the proposed method, and reconstructed with the parameters estimated using three randomly selected observation sites at two observation sites: (a) near the center of observation site area; (b) with the longest epicentral distance.\color{black}}
    \label{fig:twinexp_psd_reconst}
\end{figure*}

The relation between the reconstruction error of the seismic wavefield and the iteration number of estimation [eq.~\eqref{eq:reconst_seiserr}--\eqref{eq:reconst_paramcali}] is described in Fig.~\ref{fig:twinexp_err} in the case of three observation sites used for the model parameter estimation. The reconstruction error of the seismic wavefield was calculated as follows:
\begin{align}
\epsilon = \frac{||\mathbf{X}^{\rm true}-\mathbf{X}^{\rm reconst}||_{\rm F}}{||\mathbf{X}^{\rm true}||_{\rm F}},
\label{eq:recosterr}
\end{align}
where the notation $\|\circ\|_\mathrm{F}$ is the Frobenius norm that is the matrix norm as the square root of the sum of the absolute squares of the matrix elements. The matrix $\mathbf{X}^{\rm true}$ is the true data calculated using the true model parameters, and $\mathbf{X}^{\rm reconst}$ is the reconstructed data using the estimated model parameters based on the observed signals at three observation sites. \color{\ReviewerA}The values of the random selection shown in Fig.~\ref{fig:twinexp_err} are an average over the results using 100 patterns of subsets of three randomly selected observation sites. \color{black}
Figure~\ref{fig:twinexp_errmap} shows the reconstruction error distributions of the seismic wavefields using three observation sites selected by the proposed method and using three randomly selected observation sites, respectively. The reconstruction error presented in Fig.~\ref{fig:twinexp_errmap} is evaluated at each observation site when the number of iterations $t$ is 20. 
\color{\ReviewerA}The values for case of the random selection are the result using one pattern of subsets of three randomly selected observation sites in Fig.~\ref{fig:twinexp_errmap}. The observation sites selected using the proposed method and those randomly selected are numbered and marked with an asterisk in Fig.~\ref{fig:twinexp_errmap}, respectively. \color{black}
\color{\ReviewerA}Figure~\ref{fig:twinexp_psd_reconst} shows the power spectral density of seismic acceleration in the NS, EW, and UD directions at two observation sites, excluding the selected observation sites by the proposed method and those randomly selected in semilog plots. The two observation sites are marked by (a) and (b) in Fig.~\ref{fig:twinexp_errmap}, which correspond to the observation site near the center of the observation area and the one with the longest epicentral distance, respectively. \color{black}

\begin{figure*}
    \centering
    \includegraphics[width=13cm]{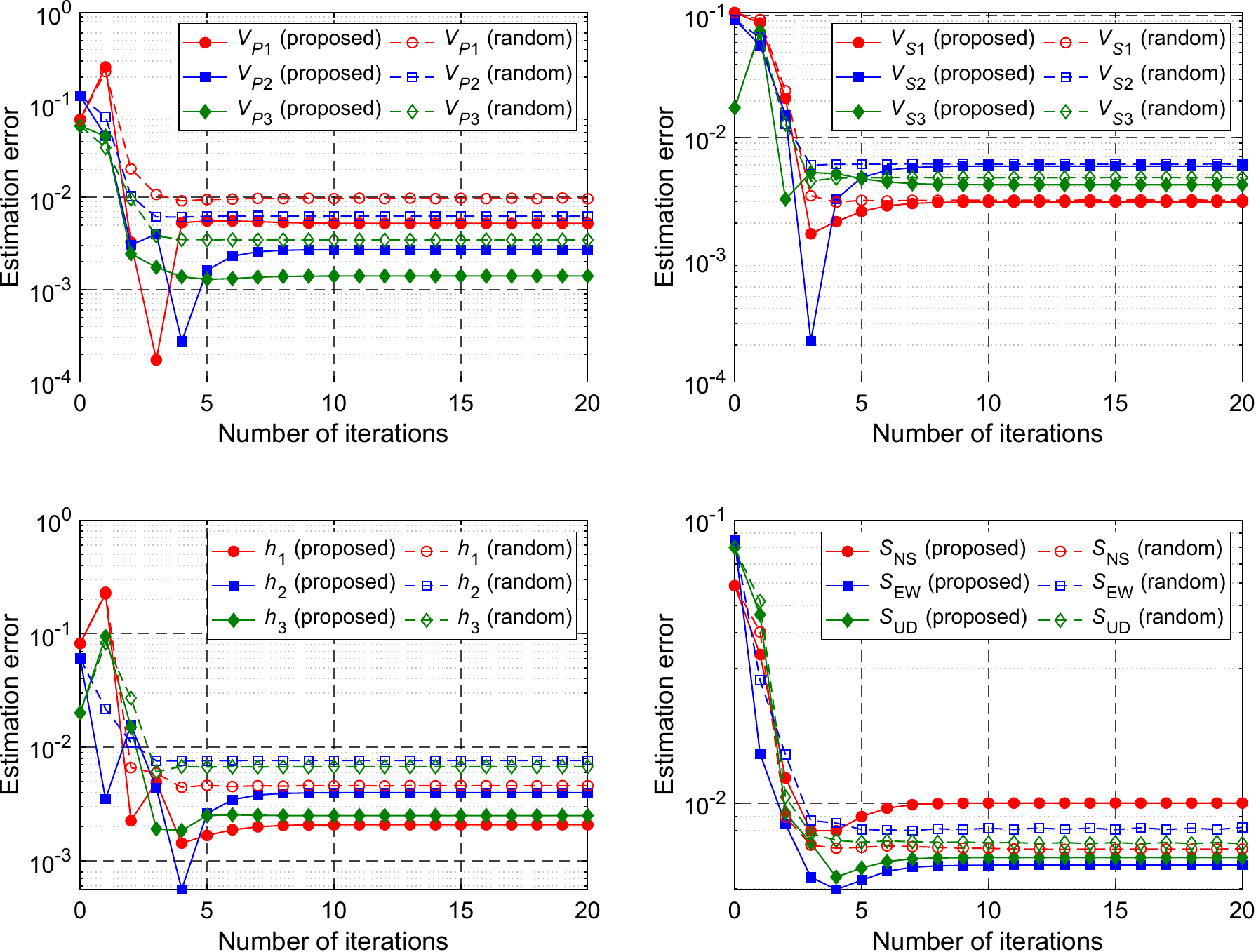}
    \caption{Comparison of the error in the model parameters estimated using the three observation sites selected by the proposed method and using three randomly selected observation sites.}
    \label{fig:twinexp_paramerr}
\end{figure*}

Figure~\ref{fig:twinexp_err} indicates that although the reconstruction error decreases as the number of iterations for optimisation of the model parameter estimation increases both in the results using the observation sites selected by the proposed method and in the results using the randomly selected observation sites, the reconstruction error can be minimized by the proposed method. 
In addition, Fig.~\ref{fig:twinexp_errmap} demonstrates that in the result using the randomly selected observation sites, although the reconstruction error is relatively small at and around the three selected observation sites, the reconstruction error is significant at the sites away from the three sites. On the other hand, the estimation using the proposed method can reduce the reconstruction error compared to that using the randomly selected observation sites at almost all observation sites. 
\color{\ReviewerA}Figure~\ref{fig:twinexp_psd_reconst} indicates that the PSD reconstructed with the parameters estimated using the observation sites selected by the proposed method matches the true values more accurately than that using the randomly selected observation sites, although the difference between the PSD reconstructed using the proposed method and that using random selection is unfortunately invisible in this range. Although these small differences are due to the assumption in the present experiment which gives the quite small discrepancies between the initial and true values of the model parameters in the present study, the estimation is somehow improved in this small range by using optimised sensor locations as previously shown in Fig.~\ref{fig:twinexp_err}. The small but certain improvement in the estimation shows the selected locations are sensitive to the parameters of the model. The sensitive sensor location is expected to be also applicable to the parameter search in the practical nonlinear problem, which is left for the future study. \color{black}

Figure~\ref{fig:twinexp_paramerr} shows the relation between the estimation error of each model parameter and the iteration number of estimation (eq.~\eqref{eq:reconst_seiserr}--\eqref{eq:reconst_paramcali}), where the estimation error of the model parameter was defined as the relative error between the estimated value and the true value. 
Although the estimation error converges to a constant value as the number of iterations increases both in the results using the observation sites selected by the proposed method and in the results using the randomly selected observation sites, the proposed method can reduce the estimation error of almost all parameters compared to that using the random selection. These results confirm that the optimised observation sites using the proposed method can reconstruct the seismic wavefield with high accuracy by estimating the model parameters accurately as expected. \color{\ReviewerA}Note that the small differences are due to the assumption in the present experiment which gives the quite small discrepancies between the initial and true values of the model parameters in the present study as stated above. \color{black}

\begin{figure*}
    \centering
    \includegraphics[width=5.5cm]{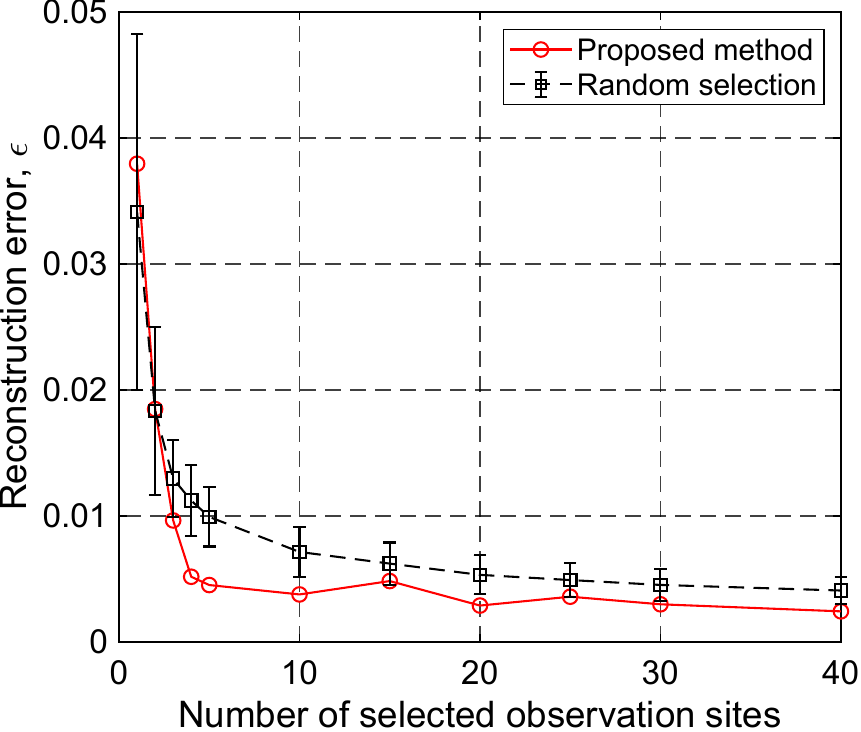}
    \caption{\color{\ReviewerA}Reconstruction error with respect to the number of selected observation sites.\color{black}}
    \label{fig:twinexp_err_obsnum}
\end{figure*}

\color{\ReviewerA}The relationship between the number of observation sites and the reconstruction error is described in Fig.\ref{fig:twinexp_err_obsnum}. Figure~\ref{fig:twinexp_err_obsnum} demonstrates that the reconstruction error tends to decrease as the number of observation sites increases. The reconstruction error with the parameters estimated using the observation sites selected by the proposed methods asymptotically approaches a small value when more than three observation sites are used. On the other hand, the error using the random selection becomes as small as that using the proposed method when 15 or more observation sites are used. Therefore, the superiority of the proposed method is remarkable for the reconstruction using 10 or less observation sites in the numerical experiments of the present study. 
However, with regard to the reconstruction using only one or two observation sites, the reconstruction error using the selected observation sites by the proposed method is within the error bars of that using the randomly selected observation sites. This is because the proposed method first selects an observation site that contributes the most to the minimization of the determinant of the normalized parameter sensitivity matrix as described in eq.~\eqref{eq:obj_detvec2}, and the strategy is not exactly consistent with selecting one that most reduces the reconstruction error of the seismic wavefield. 
On the other hand, as the number of observation sites increases, the proposed method can select a subset of observation sites suitable for accurate estimation for almost all parameters, since the determinant represents the average sensitivity of all parameters. Therefore, the superiority of the proposed method emerges when the number of observation sites is relatively small, within the range formulated in the combinatorial optimisation problem. The condition is between three and ten observation sites in the numerical experiments of the present study.
\color{black}

\section{Conclusions}
\label{sec:conclusions}
In the present study, we proposed an observation site selection method for the accurate reconstruction of the seismic wavefield by process-driven approaches. 

First, the formulation of the proposed method was presented. 
The proposed method selects observation sites suitable for accurately estimating physical model parameters to be input into a numerical simulation of the seismic wavefield. 
In the step of the observation site selection, we assume that the relation between the model parameters to be estimated and waveform data at observation site candidates is described with a linear equation. Then, the coefficient matrix represents the sensitivity of the observation site candidates to the model parameters. Therefore, the proposed method evaluates the objective function based on the D-optimality criterion of the normalized parameter sensitivity matrix in order to select the observation sites with high sensitivity to the model parameters. 
In the step of the seismic wavefield reconstruction, the difference between the observed data and the calculated data using the simulation is evaluated at only observation sites selected by the proposed method. The values of the model parameters are repeatedly estimated until the residual error of the seismic wavefield approaches to sufficiently small. Then, the reconstructed seismic wavefield can be obtained using the numerical simulation with the estimated model parameters. 

Secondly, the characteristics of observation site selection by the proposed method were investigated in numerical experiments assuming an earthquake that actually occurred in the Kanto basin and observation sites referred to the actual distribution of MeSO-net.
In the present study, the seismic velocities ($V_P$ and $V_S$) and the thickness ($h$) of each of the three layers, and source location in three directions ($S_{\mathrm{NS}}$, $S_{\mathrm{EW}}$, and $S_{\mathrm{UD}}$) are employed as the target model parameters to be estimated.
The normalized parameter sensitivity, which can evaluate the sensitivity to each parameter of each observation site, is introduced, and the proposed method is verified. 
With regard to the sensitivity to $V_P$ and $V_S$, the magnitude relationship between layers and the spatial structure is characterized by the phase shift due to the difference in the velocities and the energy distributions of the P-and S-waves. With regard to the sensitivity to $h$, the characteristics of that are determined by the sensitivity to $V_P$ and $V_S$. With regard to the sensitivity to $S_{\mathrm{NS}}$, $S_{\mathrm{EW}}$, and $S_{\mathrm{UD}}$, the factors that potentially affect the characteristics are the positional relationship between observation sites and hypocenter, and the energy distribution of the S-wave. Comparison of the normalized parameter sensitivity and the results of observation site selected by the proposed method show that the proposed method preferentially selects observation sites with high sensitivity to the parameters. 

Finally, numerical experiment of the seismic wavefield reconstruction is conducted using synthetic observation data in order to verify the proposed method. 
The reconstruction error of the seismic wavefield can be reduced using the observed data at observation sites selected by the proposed method compared to that at randomly selected observation sites. In addition, the estimation error of the parameters in the proposed method are also lower than that in the random selection. These results confirm that the proposed method can reconstruct with high accuracy by estimating the model parameters accurately.

In addition to the superiority of the reconstruction error, the proposed method can prioritise the observation sites in terms of the contribution to the accurate seismic wavefield reconstruction. The proposed method enables to make a prioritised list of observation sites according to the area, source mechanisms and earthquake types in advance. The prioritised list contributes to the accurate and quick reconstruction of the seismic wavefield when an earthquake occurs. Furthermore, the proposed method has the potential to indicate the most appropriate configuration of seismometers when they are newly installed.

The sensitivity analysis brought the physical insight into factors that affect the sensitivity to the parameters such as the seismic velocity, layer thickness, and the hypocenter location. The results of the present study were shown to be useful not only for selecting effective observation sites but also for deeply understanding the seismic phenomena. On the other hand, some assumptions were adopted in the present study. For example, although a 1-D horizontally layered subsurface structure is assumed in the present study, the nonuniformity in the horizontal direction affects the seismic wavefield in a real world. 
In future research, sensitivity analysis and observation site selection will be investigated in more realistic situations. Furthermore, the applicability of the proposed method to real observed data will be investigated. 

\section*{Acknowledgment}
This work was supported by JST CREST (JPMJCR1763).

\section*{Data Availability}
The code developed by \citet{hisada1995efficient} for the numerical calculation of the theoretical waveforms is available through the website at \url{http://kouzou.cc.kogakuin.ac.jp/Open/Green/}. 
The data of earthquake mechanism information was provided by the National Research Institute for Earth Science and Disaster Resilience at \url{https://www.fnet.bosai.go.jp/event/joho.php?LANG=en}.

\bibliographystyle{gji}
\bibliography{xaerolab}


\label{lastpage}

\end{document}